\documentclass[aps,prd,twocolumn,superscriptaddress,preprintnumbers,floatfix,nofootinbib]{revtex4-2}

\usepackage{graphicx}
\usepackage{amsmath}
\usepackage[caption=false]{subfig}
\usepackage{siunitx}
\usepackage{placeins}
\usepackage{color}
\usepackage{standalone}
\usepackage{dcolumn}
\usepackage{gensymb}
\usepackage{tensor}
\usepackage{bm}
\usepackage{microtype}
\usepackage{etoolbox}
\usepackage{amssymb}
\usepackage{mathrsfs}
\usepackage{accents}
\usepackage[normalem]{ulem}
\usepackage[dvipsnames]{xcolor}
\usepackage[colorlinks,urlcolor=NavyBlue,citecolor=NavyBlue,linkcolor=NavyBlue,pdfusetitle]{hyperref}
\usepackage[all]{hypcap}
\usepackage{enumerate}

\AtBeginDocument{\usepackage{booktabs}}

\newtoggle{commentsoff}
\ifdefined\nocomments
    \toggletrue{commentsoff}
\fi

\iftoggle{commentsoff}{
  \newcommand*{\mi}[1]{}
  \newcommand*{\sm}[1]{}
  \newcommand*{\comment}[1]{}
  \newcommand*{\warn}[1]{}
  \newcommand*{\todo}[1]{}

}{
  \newcommand*{\mi}[1]{{\color{purple}~\textsf{[\textbf{@Max}: #1]}}}
  \newcommand*{\sm}[1]{{\color{ForestGreen}~\textsf{[\textbf{@Simone}: #1]}}}
  \newcommand*{\comment}[1]{{\color{blue} \textsf{[\textbf{NOTE}: #1] }}}
  \newcommand*{\warn}[1]{{\color{red} \textsf{[\textbf{WARNING}: #1] }}}
  \newcommand*{\todo}[1]{{\color{purple} \textsf{[\textbf{TODO}: #1]}}}

}

\graphicspath{{./fig/}}

\newcommand{\ts}{\textsuperscript}

\newcommand{\beq}{\begin{equation}}
\newcommand{\eeq}{\end{equation}}

\newcommand*{\eq}[1]{Eq.\ \eqref{eq:#1}}
\newcommand*{\fig}[1]{Fig.\ \ref{fig:#1}}
\newcommand*{\sect}[1]{Sec.\ \ref{sec:#1}}

\newcommand{\infd}{{\rm d}}
\newcommand{\frot}{f_{\rm rot}}
\newcommand{\fgw}{f}

\newcommand{\data}{\mathbf{B}}

\newcommand{\dataraw}{\mathbf{d}}

\newcommand*{\hyp}[1]{{\cal H}_{\rm #1}}
\newcommand{\hypn}{\hyp{N}}
\newcommand{\hyps}{\hyp{S}}

\newcommand*{\bayes}[2]{{\cal B}^{\rm #1}_{\rm #2}}
\newcommand{\bayessn}{\bayes{S}{N}}

\newcommand{\p}{P}

\newcommand{\romer}{R\o mer}

\newcommand{\skyloc}{\hat{\Omega}}
\newcommand{\deltasky}{\Delta{\skyloc}}
\newcommand{\oorb}{\omega_{\rm orb}}
\newcommand{\osid}{\omega_\text{sid}}

\newcommand{\dcc}{LIGO-P2000413}

\graphicspath{{./fig/}}

\begin{document}

\preprint{\dcc}

\title[CW significance]{Establishing the significance of continuous\\
gravitational-wave detections from known pulsars}

\author{Maximiliano Isi}
\email[]{maxisi@mit.edu}
\thanks{NHFP Einstein fellow}
\affiliation{
LIGO Laboratory, Massachusetts Institute of Technology, Cambridge, Massachusetts 02139, USA
}%

\author{Simone Mastrogiovanni}
\email{simone.mastrogiovanni@apc.in2p3.fr}
\affiliation{AstroParticule et Cosmologie (APC), 10 Rue Alice Domon et L\'eonie Duquet, Universit\'e de Paris, CNRS, AstroParticule et Cosmologie, F-75013 Paris, France}

\author{Matthew Pitkin}
\affiliation{Department of Physics, Lancaster University, Lancaster LA1 4YB, United Kingdom.}

\author{Ornella Juliana Piccinni}
\affiliation{INFN, Sezione di Roma, I-00185 Roma, Italy.}

\hypersetup{pdfauthor={Isi, Mastrogiovanni, Pitkin, Piccinni}}

\date{\today}

\begin{abstract}
We present a method for assigning a statistical significance to detection 
candidates in targeted searches for continuous gravitational waves from 
known pulsars, without assuming the detector noise is Gaussian and stationary.
We take advantage of the expected Doppler phase modulation of the signal induced by Earth's orbital motion, as well as the amplitude modulation induced by Earth's spin, to effectively blind the search to real astrophysical signals from a given location in the sky.
We use this ``sky-shifting'' to produce a large number of noise-only data realizations to empirically estimate the background of a search and assign detection significances, in a similar fashion to the use of timeslides in searches for compact binaries.
We demonstrate the potential of this approach by means of simulated signals, as well as hardware injections into real detector data.
In a study of simulated signals in non-Gaussian noise, we find that our method outperforms another common strategy for evaluating detection significance.
We thus demonstrate that this and similar techniques have the potential to enable a first confident detection of continuous gravitational waves.
\end{abstract}

\maketitle

\section{Introduction} \label{sec:intro}

In addition to short-lived gravitational waves (GWs) from compact-binary coalescences like those observed so far \cite{gw150914,2017PhRvL.119p1101A,LIGOScientific:2018mvr,GW190425,GW190412,GW190814,GW190521g,GW190521g:imp}, ground-based detectors like the Advanced Laser Interferometer Gravitational-Wave Observatory (aLIGO) \cite{TheLIGOScientific:2014jea} and Virgo \cite{TheVirgo:2014hva}, are also expected to detect persistent, almost-monochromatic signals \cite{einsthome2016, Aasi2015, cwallsky2016, cwallskybin2014, rome2016, o1cw, o2cw}.
The primary potential source of such ``continuous waves'' (CWs) are rapidly-spinning neutron stars with an asymmetry in their moment of inertia \cite{Thorne1987}.
This includes galactic pulsars known from electromagnetic observations, which are a main target for searches for continuous signals in LIGO and Virgo data \cite{tcw2013,o1cw,Abbott:2020lqk}.
The detection of gravitational waves from any of these sources would provide a new wealth of astrophysical information, as well as invaluable opportunities to learn about fundamental physics (see, e.g., \cite{Riles:2017evm} for a recent review).

There exist a number of efforts to detect gravitational waves from  known pulsars \cite{Pitkin2017, rome2016, o1cw, PhysRevD.58.063001}.
However, an outstanding problem affecting all of these searches is the lack of a well-defined procedure  to establish the statistical significance of potential detections without making the assumption that the instrumental noise is Gaussian and (semi-)stationary.
Consequently, if evidence for a continuous wave from a known pulsar was found today, we would be unable to establish, with certainty, the probability for this to have arisen from a spurious noise artifact.
The need for a systematic and robust way of computing detection significance in the presence of non-Gaussian noise has already become apparent with the appearance of hard-to-diagnose outliers in recent searches in actual aLIGO data \cite{o1cw,Abbott:2017pqa}.

Establishing a robust procedure to assign significance is challenging because the noise artifacts that limit the searches are intrinsically unpredictable and cannot be modeled from first principles.
Given this, we may instead attempt to empirically determine the response of the different searches to real detector noise in the {\em absence} of astrophysical signals.
Armed with such knowledge, we would then be able to analyze actual data, or ``foreground'', and produce empirical likelihood ratios (or other measures of detection confidence, like $p$-values) for the presence of an astrophysical signal vs just instrumental noise, Gaussian or otherwise.
This requires several instances of ``background''---that is, instrumental data that are known to contain no astrophysical signals, while still retaining all statistical properties representative of real instrumental noise.

Ideally, one would obtain background distributions by physically isolating the instruments from the environment to shield them from actual signals.
Because this is impossible in the case of gravitational waves, we must attempt to replicate this shielding digitally after the data have been recorded. 
Several techniques exist to do this when looking for gravitational-wave transients, the most straightforward of which is probably the use of ``time slides'': the outputs of different detectors are shifted relative to each other by time offsets longer than the light-travel time between them \cite{Usman:2015kfa,Canton:2014ena}.
This ensures the spuriousness of any signal candidate left in the multi-detector data thus produced, hence allowing us to estimate how likely it is for noise to mimic a signal.

The direct analog of time slides in the context of continuous waves would be ``frequency slides'': a misalignment of the frequency-domain data of different detectors. 
However, our ability to effect such frequency shifts is limited by the frequency resolution of the searches (of the order of inverse observation time), and the fact that the properties of actual instrumental noise are heavily dependent on frequency---not only due to a frequency-dependent power spectral density, but also to varying populations of narrow-band noise features.
By the same token, time slides themselves would not be feasible in transient analyses if the noise properties of the detectors changed rapidly compared to the duration of a typical signal.

In light of this, here we propose a simple method for estimating the background of searches for continuous gravitational waves by analyzing data assuming an incorrect sky location for the targeted source.
This ``sky-shifting'' takes advantage of the expected Doppler modulation of the signal due to the relative motion of detector and source to effectively blind the search to real astrophysical signals.
We can use this to produce a large number of noise-only instantiations of data, so as to empirically estimate the background of a search and assign detection significances in the presence of actual detector noise.
We demonstrate that this method can outperform another common strategy for estimating the background in realistic situations.

We begin by providing relevant background about continuous waves and targeted searches in \sect{background}.
We then introduce the sky-shifting method and explore its applicability in \sect{method}.
We demonstrate the efficacy of the strategy in \sect{results} with the aid of several examples based on both synthetic and actual detector noise.
We conclude in \sect{conclusion}.

\section{Background} \label{sec:background}

In this section, we review the basic morphology of continuous gravitational waves as measured by differential-arm detectors, with an emphasis on the timing corrections on which we will rely for sky-shifting (\sect{morphology}).
We also make a special point of discussing the relation between the frequency resolution at which a signal is sampled and the ability to localize the source in the sky (\sect{resolution}).
We next describe the key properties of noise in existing ground-based instruments as it pertains to searches for persistent signals (\sect{noise}).
Finally, we provide an overview of three staple search methods for these signals in LIGO and Virgo data (\sect{searches}): the Bayesian time-domain method, and the frequentist 5-vector and ${\cal F}$--statistic methods.

\subsection{Continuous waves} \label{sec:cws}

\subsubsection{Morphology} \label{sec:morphology}

Continuous waves are nearly monochromatic gravitational perturbations with constant intrinsic amplitude that are expected to be sourced by some rapidly spinning bodies, like neutron stars. 
Within the context of standard physics, there are several ways in which a neutron star could emit CWs, but the most favored is the presence of a nonaxisymmetry in the star's moment of inertia \cite{Jones2002}.
For this type of {\em triaxial}, nonprecessing source, such a GW will induce a strain in a differential-arm (quadrupolar) detector, like LIGO or Virgo, which can be written as:
\begin{align} \label{eq:cw}
h(t) &= h_0 \frac{1}{2}(1+\cos^2 \iota)F_+(t; \psi) \cos \phi(t) \nonumber \\
&+ h_0 \cos \iota F_\times(t; \psi)\sin \phi(t)\, ,
\end{align}
where the $F_+(t; \psi)$ and $F_\times(t; \psi)$ factors respectively give the instrument's response to the plus ($+$) and cross ($\times$) GW polarizations, $\iota$ is the inclination angle between the spin axis of the source and the observer's line-of-sight, $\phi(t)$ is the phase of the signal, and $h_0$ is an overall amplitude related to the properties of the source by:
\beq \label{eq:h0}
h_0 = \frac{16 \pi^2 G}{c^4} \frac{\epsilon I_{\text{zz}}\frot^2}{r}\, ,
\eeq
where $r$ is the source distance, $\frot$ its rotation frequency around the principal axis $z$, $\mathbf{I}$ the moment-of-inertia tensor and $\epsilon\equiv (I_{xx}-I_{yy})/I_{zz}$ the equatorial ellipticity \cite{Thorne1987}.

The antenna patterns, $F_+(t;\psi)$ and $F_\times(t;\psi)$, encode the amplitude modulation of the signal due to the local geometric effect of a GW acting on a given detector.
Thus, they implicitly depend on the relative location and orientation of source and detector by means of the source's right-ascension $\alpha$, declination $\delta$, and polarization angle $\psi$.
The latter gives the orientation of the frame in which the polarizations are defined, and we set it to be the angle between the line of nodes and the projection of the celestial North onto the plane of the sky.
While $\alpha$ and $\delta$ are always well known, $\psi$ generally is not, which is why we show this argument explicitly.
Importantly, the antenna patterns acquire their time dependence from the rotation of Earth on its axis, and consequently have a characteristic period of a sidereal day (${\sim}10^{-5}$ Hz).

For a simple triaxial source, the GW frequency $\fgw$ is twice the rotational value $\frot$, so we can write:
\beq\label{eq:phase}
\phi(t)=2\phi_{\rm rot} (t)+\phi_0,
\eeq
where $\phi_{\rm rot}$ is the rotational phase as measured via electromagnetic (EM) observations and $\phi_0$ is a fiducial phase offset.
The rotational frequency itself is almost constant, with a small spin-down due to energy loss into the environment (via GWs and other mechanisms), which means that the phase evolution can be well described by a simple Taylor expansion on $\tau$, the time measured by a clock inertial with respect to the source:
\beq
\phi(t) = 2\pi \sum^N_{j=0} \frac{\partial_t^{(j)} f_0}{(j+1)!} \left[\tau(t) - T_0 \right]^{(j+1)} .
\eeq
Here $\partial_t^{(j)} f_0$ is the $j$\ts{th} time derivative of the GW frequency measured at the fiducial time $T_0$, and $N$ is the order of
the series expansion (1 or 2 suffices for most sources).
Timing solutions are generally obtained through the pulsar timing package TEMPO2 \cite{Edwards2006}. 
These solutions have exquisite precision (frequency uncertainty of $10^{-12} \lesssim \delta \frot \lesssim 10^{-8}$ Hz for most pulsars) and are the cornerstone of targeted searches for continuous waves from known pulsars.

The inertial time, $\tau$ in \eq{phase}, is usually taken to be the time measured by a clock at the Solar System barycenter (SSB), which is itself assumed to be inertial with respect to the pulsar.
In that case, $\tau$ can be written as a function of detector time, $t$, by taking into account some well-known, time-dependent offsets:
\beq \label{eq:time}
\tau (t) = t + \Delta_{\rm E}(t) + \Delta_{\rm S}(t) + \Delta_{\rm binary}(t) + \Delta_{\rm R}(t)\, . 
\eeq
Here $\Delta_{\rm E}$ is the Solar-System Einstein delay; $\Delta_{\rm S}$ is the Solar-System Shapiro delay; $\Delta_{\rm binary}$ is the delay originating from the motion of the pulsar in its binary (a term that vanishes for isolated sources) \cite{Dupuis2005}; and $\Delta_{\rm R}$ is the kinematic delay due to the relative motion of the detector with respect to the source.

The timing correction of \eq{time} is heavily dependent on the sky-location of the targeted pulsar and will be the key to the sky-shifting method presented in \sect{method}.
The dependence on sky location is dominated by the last term in \eq{time}, $\Delta_{\rm R}$.
This is sometimes known as the ``\romer{} delay'' and encodes the Doppler modulation of the signal:
\beq \label{eq:romer}
\Delta_{\rm R} (t) = - \frac{\skyloc \cdot \vec{r}(t)}{c}\, ,
\eeq
where $\vec{r}(t)$ is a vector joining the SSB and the detector at any given time, $\skyloc$ is a unit vector pointing from the SSB in the direction of the source,%
\footnote{The source-location vector, $\skyloc$, can be treated as constant over the timescale of our observations.}
and $c$ is the GW speed.
For practical purposes, $\vec{r}$ is usually computed by first splitting it into three components:
\beq
\vec{r} = \vec{r}_\odot + \vec{r}_\oplus + \vec{R}\, ,
\eeq
with $\vec{r}_\odot$ joining the SSB with the Sun's center, $\vec{r}_\oplus$ joining Sun and Earth, and $\vec{R}$ going from the Earth's center to the detector on the surface.
One can then use Solar System ephemerides, together with knowledge of the location of the detector on Earth and the source in the sky, to compute the \romer{} correction at any given time.

The timing correction of \eq{time} can be understood as inducing extrinsic frequency shifts to the signal, as seen by the detector.
This is dominated by the \romer{} term, $\Delta_R(t)$, which results in a modulation at the frequency of Earth's orbital rotation, $\oorb \approx 2\times 10^{-7}\, \mathrm{rad/s}$, as well as subdominant daily effects due to its spin, $\osid=2\pi/\text{(sidereal day)}\approx7\times10^{-5}\, \mathrm{rad/s}$.
In the frequency domain, the effect of this correction is to spread the signal power across a narrow band centered on its intrinsic GW frequency, with a characteristic width of $\Delta f/ \fgw \approx \oorb r_\oplus / c \approx 10^{-4}$ (see e.g.~\cite{Sammut:2013oba}).
This frequency modulation will be the key to our approach.

\subsubsection{Frequency and sky resolution} \label{sec:resolution}

The sky resolution is the minimum angular separation in the sky at which two, otherwise equal, sources could be distinguished.
This is a function of the frequency resolution at which the signal is sampled, namely:
\beq
\delta f = 1 / T\, ,
\eeq
for an observation time $T$.
This frequency bin is related to a sky-bin by the sky-location--dependent frequency modulation of \eq{time}.
Thus, the angular resolution will be roughly given by the separation in the sky corresponding to a \romer{} frequency shift of $\Delta f = \delta f$.
In other words, we may define a bin around any point in the sky by the maximum angular distance one can move away from that point before the frequency shift caused by the modulation of \eq{romer} reaches a magnitude of $1/T$.
Thus, the characteristic size of a bin defined this way will necessarily depend on the integration time.

\begin{figure*}
\includegraphics[scale=0.45]{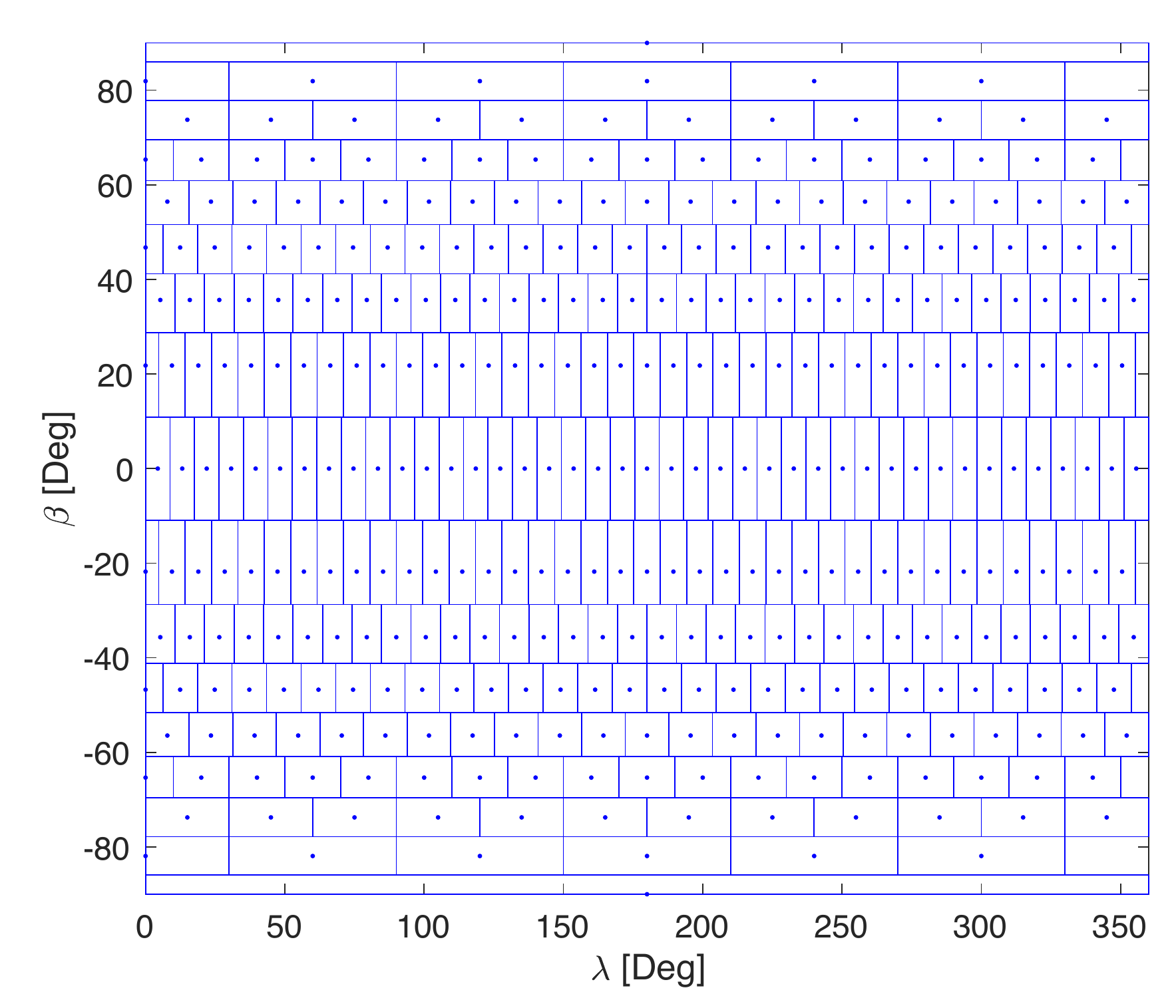}
\includegraphics[scale=0.45]{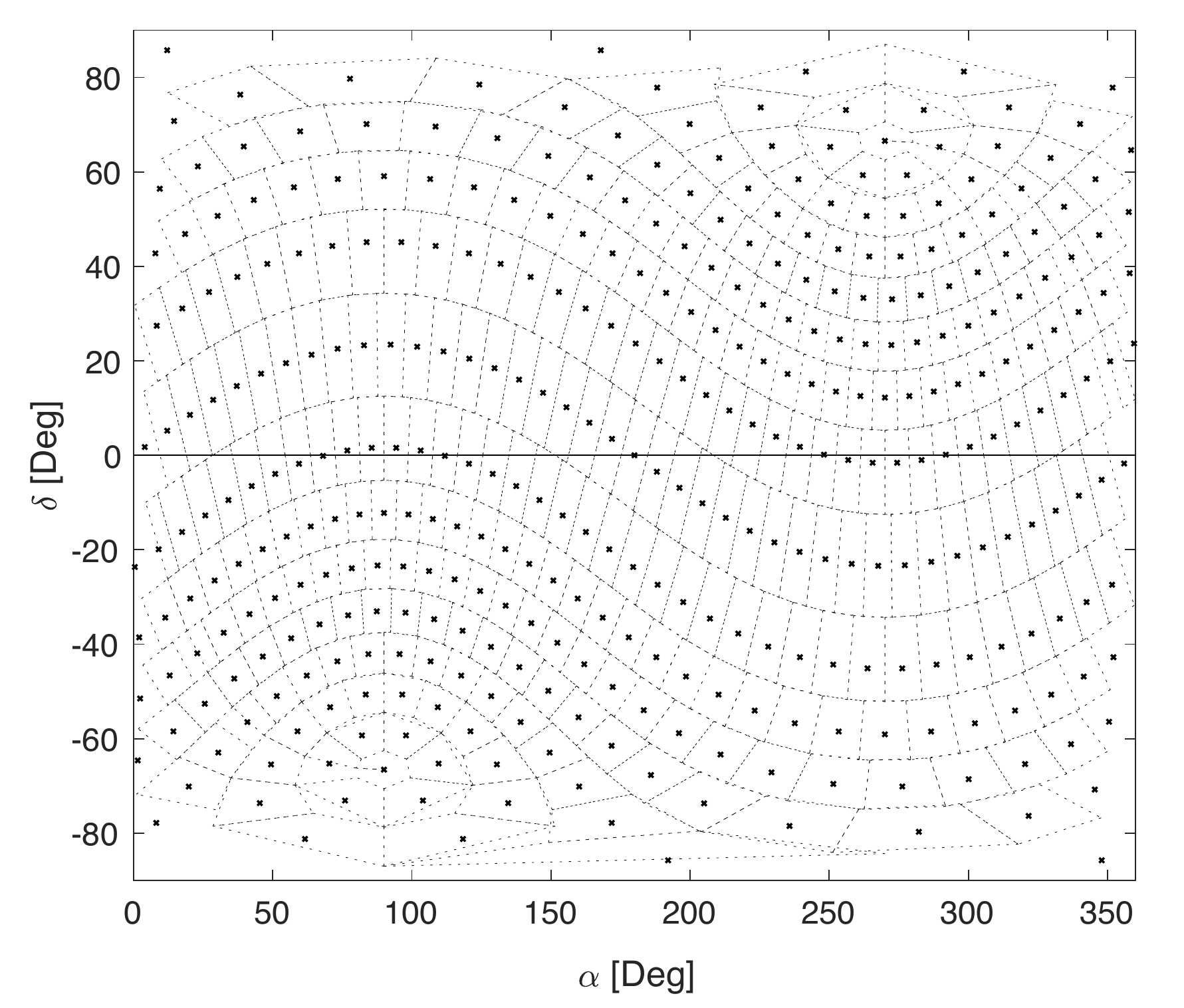}
\caption[Sky-grid example in equatorial and ecliptic frames]{{\it Left:} Sky bins built using ecliptic coordinates for a search at a frequency of 60 Hz and an integration time of 1024 s. {\it Right:} The same sky grid for the same sky configuration in equatorial coordinates.}
\label{fig:coord}
\end{figure*}

Proceeding as above, we may cover the sky with a series of such bins to obtain a ``sky grid'' representing the resolvability of points in the sky as a function of angular location.
Because the timing correction of \eq{time} is dominated by $\Delta_R$, which is itself mostly due to Earth's \emph{orbital} motion, such a sky grid will be most naturally defined in \emph{ecliptic} coordinates to yield bin sizes given approximately by \cite{method_allsky}
\begin{align}
& \delta \beta = \frac{1}{N_d |\sin \beta|}\quad, \quad \delta \lambda= \frac{1}{N_d |\cos \beta|}
\label{eq:skybin}
\end{align}
where $\beta$ and $\lambda$ are respectively the ecliptic latitude and longitude, and the scale factor (that represents the total number of Doppler bins) is
\beq \label{eq:skybin_scale}
N_d= \Delta f \cdot f \cdot T = \frac{\fgw \, \oorb r_\oplus T}{c}
\eeq
for Earth's orbital radius $r_\oplus$.
As demonstrated in \fig{coord}, the sky-grid can be easily computed using ecliptic coordinates (left panel) and then rotated to equatorial coordinates (right panel).
This is a conservative sky-grid for an integration time of $T_{\rm coh}=1024 s$ that implicitly assumes the power of the signal may be split over at most two frequency bins as a result of the timing correction---in practice, the characteristic size of the sky bins may be reduced, but the scaling with $\fgw$ and $T$ will always be as in \eq{skybin_scale}. In a full-coherent search that uses $1$ year of data, the sky-bins size are significantly reduced.

\subsection{Detector noise} \label{sec:noise}

The output of ground-based gravitational-wave detectors is vastly dominated by instrumental noise \cite{TheLIGOScientific:2016agk,Martynov:2016fzi}.
For this reason, the weak continuous signals discussed in \sect{cws} are expected to become visible only after long periods of coherently-integrated observation.
Understanding the statistical properties of the noise is critical to successfully detecting these signals. %

For the most part, the noise in a given detector is well described as a Gaussian random process with a frequency-dependent (colored) power spectral density  \cite{TheLIGOScientific:2016agk,Martynov:2016fzi}.
Gaussian noise has numerous convenient statistical properties that would drastically simplify many of LIGO and Virgo's analyses.
However, this idealization is far from perfect: the data are plagued with uncountable non-Gaussian features with a range of spectral properties and durations.
Among these, the most-often discussed are probably the noise transients (``glitches'') that haunt searches for compact-binary coalescences \cite{2017arXiv171002185T}.
Yet, searches for continuous waves are most affected not by these short-lived glitches, but rather by persistent narrow-band features (``lines'') \cite{Covas:2018oik}.
Many of these spectral lines only become apparent after long-periods of coherent observation, making their identification and eradication especially challenging.
Noise-spectral lines could also be accompanied by side-bands which affects the sensitivity in a narrow-frequency region \cite{2018PhRvD..97h2002C}.
Furthermore, their distribution over the sensitive frequency band of the detectors is highly irregular and changes with time as the instruments evolve.

Fully-coherent searches for continuous waves tend to have very high frequency resolution (of order $\delta f \sim 10^{-7}$--$10^{-8}$ Hz), scaling directly with the integration time ($\delta f \sim 1 / T$).
This fine resolution means that such analyses can fall victim to very narrow (and weak) noise lines.
Furthermore, as mentioned at the end of \sect{morphology}, a pulsar signal will be spread over a band of width $\Delta f \approx \fgw \cdot 10^{-4}$ Hz around its central GW frequency $\fgw$.
This means that attempts to find such a signal will be affected by noise over a range of frequencies, broad with respect to the typical resolution of the search.
A persistent departure from Gaussianity in that frequency range (e.g.~a wandering instrumental line that happens to intermittently cross the targeted band) will confound most searches, potentially yielding false positives (``outliers'').
Naturally, the number of outliers due to unmodeled noise found by the pipelines will increase with the searched CW parameter space, as well as with observation time (which increases the frequency resolution and the sensitivity to unmodeled noise sources).

As is the case with the glitches affecting searches for compact binaries, lines and other non-Gaussian features would not be an issue for continuous-wave searches if there existed a robust way to model them and directly incorporate that knowledge into the statistical analyses (cf.~\sect{searches} below).
However, the noise artifacts in the set that interests us are, by definition, impossible to fully model from first principles: any particular noise source that {\em is} well-understood can usually be physically or digitally removed, so that they are no longer of concern \cite{Leaci:2010zz,Meadors:2013lja,Driggers:2018gii,Covas:2018oik}; the remaining artifacts are, therefore, those that are intrinsically unpredictable or so-far not understood.
Consequently, we are left to try to find ways to {\em empirically} estimate the true statistical background (i.e.~the probability distribution of false-positives) of a search in order to assign significances to potential detection candidates.

\subsection{Searches} \label{sec:searches}

Searches targeted at known pulsars make use of the simple form of the expected signal, \eq{cw}, to match-filter the data and determine the likelihood that a signal is present.
There exist three standard approaches of this kind: the time-domain Bayesian \cite{Niebauer1993, Dupuis2005, Pitkin2006, Pitkin2017} and ${\cal F}/{\cal G}$-statistic \cite{PhysRevD.58.063001,Jaranowski2010,2014PhRvD..89f4023K,2015CQGra..32c5004K} methods, and the frequency-domain $5n$-vector method \cite{Astone2010, Astone2012, Astone2014}.
Due to the technical details underlying each implementation, only the Bayesian time-domain method has been broadly applied to a large number of targets \cite{o1cw}.
Although sky-shifting is applicable to all three of these techniques, in the following sections we will only use the Bayesian and 5-vector searches for concrete examples.
For completeness, here we provide a brief overview of the basics of all three approaches.

\subsubsection{Bayesian approach} \label{sec:searches_bayes}

Bayesian statistics provide a complete and straightforward framework for computing the probability that a given set of data contain a signal vs Gaussian noise, and for inferring the parameters that best describe the signal if present.
The implementation utilized for searches by the LIGO Scientific Collaboration \& Virgo Collaboration \cite{Pitkin2017} takes advantage of the fact that the phase evolution $\phi(t)$ is known from electromagnetic observations to remove the high-frequency components of the signal early in the process---this dramatically simplifies the Bayesian inference step itself \cite{Dupuis2005, Pitkin2006}.

First, the data are digitally heterodyned \cite{Niebauer1993, Dupuis2005}, so that the signal they putatively contain becomes:
\beq \label{eq:het-data} 
h'(t)\equiv h(t) e^{-i\phi(t)} =\Lambda(t) + \Lambda^*(t) e^{-i2\phi(t)}\, , 
\eeq
with $*$ indicating complex conjugation, and
\beq \label{eq:het-signal}
\Lambda(t) \equiv \frac{1}{4} F_+(t) h_0 (1+\cos^2 \iota) - 
\frac{i}{2} F_\times(t) h_0 \cos\iota\, .
\eeq
A series of low-pass filters are then applied to remove the second term in \eq{het-data}, which enables the down-sampling of the data by averaging over minute-long time bins.
As a result, $\Lambda(t)$ is the only contribution from the original signal left in our binned data, $\data$, which will now look like
\beq \label{eq:data}
B_{\rm expected}(t_{k})= \Lambda(t_k) + n(t_{k}),
\eeq
where $n(t_{k})$ is the heterodyned, filtered and downsampled noise in bin $k$, which carries no information about the GW signal.

\eq{data} implies that the quantity $B(t_k) - \Lambda (t_k)$ should have the statistical properties of noise, and that \eq{het-signal} should be the template in our search.
This knowledge can be used to compute the marginalized-likelihood ratio (Bayes factor) that the data contain a signal buried in noise ($\hyps$), vs just Gaussian noise ($\hypn$):
\beq \label{eq:bsn}
\bayessn = \frac{\p(\data \mid \hyps)}{\p(\data \mid \hypn)}\, .
\eeq
If the detector noise was indeed Gaussian, this single quantity would suffice to define a detection criterion: a value greater than unity would indicate the signal model is favored by that factor (in terms of betting odds), and {\em vice versa}.
However, since actual noise cannot be guaranteed to be Gaussian (and, generally, will not be), the probability ratio of \eq{bsn} does not inform us about the relative likelihoods of a signal vs \emph{actual} (non-Gaussian) noise.
To address this, one may attempt to capture instrumental artifacts by defining a construction similar to \eq{bsn} but using signal-coherence across detectors to distinguish spurious effects from actual astrophysical signals \cite{Keitel2014, Isi2017, o1cw}.
Nevertheless, it cannot be shown that any such construction will always capture all the features of real instrumental noise (in the language of formal logic, our hypothesis set is never complete).
Therefore, we would benefit from a method to empirically test the efficacy of our Bayesian constructions at actually distinguishing signals from (non-Gaussian) detector noise.

\subsubsection{5-vector approach} \label{sec:searches_5vector}

The frequentist 5-vector method \cite{Astone2010} builds a detection statistic using the sidereal modulation given by the interferometer antenna response to the two CW polarizations, encoded by $F_{+/\times}$ in \eq{cw}.
Similarly to the procedure outlined in \sect{searches_bayes}, the first step is to remove all the possible phase modulations, apart from the sidereal ones caused by the antenna patterns.
Depending on the type of search, this may be achieved through different techniques, including subheterodyning, nonuniform resampling, or a combination thereof \cite{2017CQGra..34m5007M, 2014PhRvD..89f2008A,2019CQGra..36a5008P}.
After this step, the signal can be modeled via two sidereal responses,  $A_{+/ \times}(t)$ , analogous to $F_{+/\times}(t)$ but which do not depend on the polarization angle $\psi$ (see \cite{Astone2010} for more details).
Concretely, it may be shown that $A_{+}(t) \equiv F_{+}(t; \psi=0)$ and $A_{\times}(t) \equiv F_{\times}(t; \psi=\pi/4)$.
Doing this, the signal assumes the complex-valued form:
\beq \label{eq:5vec}
h(t)=H_0 (\eta) \left[ H_+ (\psi, \eta) A_+ (t) + H_\times (\psi, \eta) A_\times (t) \right] , 
\eeq
where $\eta$ is related to the ratio of the two polarization amplitudes given in \eq{cw},
\beq
\eta=-\frac{2 \cos \iota}{1+\cos^2 \iota^2}
\eeq
and with $H_{+/\times}$ defined by
\begin{eqnarray*}
H_+ =\frac{\cos(2 \psi) - i \eta \sin (2 \psi)}{\sqrt{1+\eta^2}}\, , \\
H_\times =\frac{\sin(2 \psi) - i \eta \cos (2 \psi)}{\sqrt{1+\eta^2}}\, .  
\end{eqnarray*}
Just as in the case of $\Lambda(t)$ in \eq{het-signal}, the frequency components of a signal described by \eq{5vec} are simply those corresponding to the sidereal modulations encoded in $A_{+/ \times}(t)$.
These frequency components ($f^i_\text{5-vec}$) are integer multiples of the sidereal rotation frequency of Earth, namely:
\beq \label{eq:5vec_basis}
f^i_\text{5-vec}=f_{\rm gw}+2 \pi k^i \, \osid~,~\vec{k}=\left[-2,-1,0,1,2\right] .
\eeq
Therefore, any signal like \eq{5vec} may be described as a vector in the space spanned by the five $\delta$-functions corresponding to the frequencies in \eq{5vec_basis}.

To search for signals, the frequency domain GW data can be projected onto the 5-vector space, to obtain a set of projections $\vec{X}$.
This resulting vector now lives in the same space as the sidereal templates, which can be represented as 5-vectors $\vec{A}_{+ / \times}$.
We may thus obtain the matched-filter between the data and the antenna patterns by taking a simple scalar product between $\vec{X}$ and  $\vec{A}_{+ / \times}$. 
By maximizing this matched-filter, one obtains an estimator for the GW polarization amplitudes:
\beq
\widehat{H}_{+/ \times}= \frac{\vec{X} \cdot \vec{A}_{+ / \times}}{|\vec{A}_{+ / \times}|^2}\, ,
\eeq
which can be in turn used to define a detection statistic:
\beq \label{eq:5ds}
S_5 \equiv |\vec{A}_{+}|^4 |H_{+}|^2 + |\vec{A}_{\times}|^4 |H_{\times}|^2\, .
\eeq

After carrying out the above procedure for templates corresponding to different parameters, detection candidates (i.e.~values of the parameters for which might match a potential signal) are identified by their value of $S_5$.
In particular, the statistic is required to exceed a threshold corresponding to a preset false alarm probability.
To do this, one must know or measure the distribution of $S_5$ over noise.
Traditionally this has been computed analytically by assuming purely Gaussian noise with known variance \cite{Astone2010}.

Alternatively, since real data are not Gaussian, one may try to approximate the background distribution by computing $S_5$ over frequency bands far from the expected signal (``off-frequency'' analysis).
The frequency regions should be far enough from a possible CW signal such that only the noise contribution is present in the detection statistic, and close enough to the analyzed band to share its statistical properties.
Given that the noise is strongly frequency dependent, finding this sweetspot is far from trivial (if at all possible) and one can never guarantee that the conditions required for an unbiased estimation of the background are being satisfied.

\subsubsection{${\cal F}$-statistic} \label{searches_fstat}

The ${\cal F}$-statistic was first introduced in \cite{PhysRevD.58.063001} for gravitational-waves searches form neutron stars, and was later extended for other astrophysical objects (e.g., \cite{2007CQGra..24S.565P, 2012PhRvD..86l3010K}).
In the case of Gaussian noise, the ${\cal F}$-statistic is defined as the natural logarithm of the maximum-likelihood ratio between the signal and noise hypotheses:
\beq \label{eq:fstat}
{\cal F} =\max\left[ \ln \frac{\p(\dataraw \mid \vec{\theta}, \hyps)}{\p(\dataraw \mid \hypn)}\right]_{\vec{\theta}}\, , 
\eeq
where $\dataraw$ are usually calibrated detector data, and the maximization is over the signal-template parameters, $\vec{\theta}$.
 It can be shown that the ${\cal F}$ statistic can be analytically maximized over the ``extrinsic parameters'' ($h_0$, $\psi$, $\iota$ and $\phi_0$), thus reducing the dimensionality of the numerical computations to the so-called ``intrinsic parameters'' ($\alpha$, $\delta$, $\fgw$ and $\dot{f}$).
 Since in a targeted search the intrinsic parameters are supposed to be perfectly known, a targeted search based on the ${\cal F}$-statistic would reduce to the computation of one value for ${\cal F}$, that is later compared to the expected noise-only distribution for Gaussian noise in order to assign $p$-value.

The analysis proceeds by match-filtering the data against four different templates, each of them corresponding to a particular combination of intrinsic phase evolution and sidereal modulation.
The outcome of these four filters is the $\cal{F}$-statistic, for which, if the data are composed purely of Gaussian noise, it can be shown that the value $2\cal{F}$ follows a $\chi^2$-distribution with 4 degrees of freedom \cite{PhysRevD.58.063001}.
Detection candidates (``outliers'') are selected according to their false alarm probability, which can be computed analytically if one assumes Gaussian noise.
However, false-positive outliers arise when the noise is not Gaussian, and thus not properly handled by \eq{fstat}.
If this is the case and one cannot trust the background distribution of the statistic to be simple $\chi^2$, this distribution must be estimated empirically by producing sets of data known with certainty to contain no astrophysical signals.

\section{Method} \label{sec:method}

Having reviewed the basics behind targeted searches for continuous waves, including the difficulties inherent to non-idealized instrumental noise, we here introduce sky-shifting as way to empirically assign detection significances.
In \sect{offsourcing} we describe the basic ideas behind this simple procedure and explain how it can be easily applied to the Bayesian and 5-vector searches. 
In \sect{blinding}, we heuristically explore the limits of applicability of this technique, concluding that sky-shifting is a viable method for estimating the background distribution of detection statistics, as long as a few simple conditions are satisfied.
This will be demonstrated in the following section (\sect{results}) with concrete examples.

\subsection{Off-sourcing} \label{sec:offsourcing}

Lacking a satisfactory way to model all noise artifacts and their effect on CW searches from first principles, we may instead attempt to empirically determine the distribution of the different search statistics in response to real detector noise and in the absence of astrophysical signals.
As discussed in \sect{intro}, a naive attempt at blinding the data to astrophysical CWs using methods analogous to those used for CBCs is doomed to failure.
Therefore, we may instead look for a solution in specific properties exclusive to real gravitational signals, as opposed to noise.

One example of such a feature is the requirement of consistency between the phase evolution observed by EM astronomers, and the sky location of the source: while the two properties, as encoded in the signal itself, must necessarily agree in the case of a real GW, there is no special link between them in the case of noise artifacts.
Furthermore, as explained in Sec.~\ref{sec:cws}, the location of the source is independently imprinted in the morphology of the signal twice: in the amplitude modulation due to the antenna patterns, \eq{cw}, and in the frequency modulation due to \romer{} and other timing delays, \eq{time}.
Since these three properties (frequency, amplitude modulation, and phase modulation) must all agree for an astrophysical signal, we may ask: how likely is it for an instrumental artifact to randomly satisfy this condition and thus mimic a real signal from a given source?

{\em A priori}, an instrumental artifact with frequency close to that expected from a given source is no more likely to also show the amplitude and phase modulations corresponding to the true location of the pulsar than those of any other arbitrary sky location.
In other words, there is no reason for instrumental noise at the target frequency to ``know'' what the true sky location of the source is.
\emph{By carrying out our analysis assuming incorrect sky locations, we may blind ourselves to astrophysical signals and empirically estimate the probability that instrumental artifacts in the narrow frequency region corresponding to a given source also present the modulation matching its location in the sky.}

The above idea may be rephrased in the language of function spaces.
Continuous signals observed for a finite period of time can be represented as vectors in the space of square-integrable functions ($L^2$) or, after discretization, the space of square-summable sequences ($\ell^2$).
We would like to estimate the overlap between the subspace of $L^2$ (or, rather, $\ell^2$) occupied by noise features and the much narrower one spanned by the signal template, \eq{cw}.
We attempt to empirically achieve this by computing an inner product (defined by the detection statistic itself) between the data and signal templates (basis elements) corresponding to different sky locations.
We expect this to work partly because templates for different sky locations will be morphologically very similar to the true template, while the same is not true for any arbitrary function.
This also allows us to explore the statistical properties of the noise in the same region of frequency space occupied by the expected signal.
(See also \cite{Mastrogiovanni:2018mih}.)

An alternative to sky-shifting would be to simply randomly shuffle the time series data used in the analysis; this would be easy to do for the time-domain data $\data$ used in the Bayesian analysis, and is similar to previously implemented strategies (e.g.~\cite{Isi:2015cva}).
Such a shuffling would completely decohere  any real signal in the data and, given the huge number of permutations of the shuffling, any random shuffling would be extremely unlikely to be correlated with any other.
This approach would be the simplest way to proceed if the data were purely Gaussian and stationary across an observation run.
However, for real data that contains contaminating instrumental lines and a time-varying noise level, it would also scramble these components.
Therefore, any statistic produced with the scrambled data would not be a fair comparison to the unshuffled foreground.
Shuffling would also be complicated to perform for a search method that begins with data that has already been transformed into the frequency domain.
Off-sourcing, as described above, does not suffer from these limitations.

\subsubsection*{Implementation examples}

Background distributions may be estimated via sky-shifting in the context of any of the searches described in \sect{searches}.
This is true regardless of whether the search is carried out in the time or frequency domains, for one or several detectors, maximizing or marginalizing over nuisance parameters.
This generality stems from the fact that sky-shifting is largely insensitive to the specific details behind the computation, as long as the sky location and phase evolution are assumed to be known.

Let us first illustrate this by using the time-domain Bayesian search as a concrete example.
As outlined in \sect{searches_bayes}, this approach is split into two stages: ({\em i}) heterodyning of the data to put the signal in the shape of \eq{het-signal}; and ({\em ii}) Bayesian inference to compute the relative likelihood of a signal being present, \eq{bsn}.
It is important for our purposes that information about the location of the pulsar is only needed in the first step, making it straightforward to apply our suggested strategy. 
In particular, we may intentionally heterodyne the data assuming an off-source (shifted) sky location, and then carry out the inference stage as usual, assuming the true (on-source) sky location.
Rather than being indicative of a signal, a large Bayes factor obtained this way would necessarily reveal the presence of a noise artifact.
This process may be repeated for different sky locations to obtain an estimate of how likely noise is to mimic a signal from this pulsar.

As another example, consider the frequency-domain 5-vector approach of \sect{searches_5vector}.
In that case, we may also resample or reheterodyne the data assuming a shifted sky-location during a preprocessing stage. 
This procedure is expected to spread the power of a possible GW signal over many different frequency bins, making it too weak to be detectable and thus blinding the analysis to it.
Next, we compute the $S_5$ statistic by using the 5-vector sidereal function $A_{+/\times}(t)$ computed for the on-source sky position.
We can then repeat these steps for many different sky locations to obtain a collection of background values for the $S_5$ statistic.
This yields a noise-only distribution for the $S_5$ statistic that quantify the probability for a noise disturbance to mimic the sidereal antenna patterns corresponding to the true sky location.

\subsection{Blinding and draw-independence} \label{sec:blinding}

In order for the method above to work, we need to make sure that the different sky locations used are actually distinct, so that there is no leakage of a possible GW signal in the draws used fro building the noise-only statistic distribution.
In order for the method above to work, we need to make sure that: ({\em i}) the different sky locations used are actually distinct, so that the results can be treated as distinct draws from the probability distribution we are trying to estimate; and ({\em ii}) sky-shifting really does blind the data to foreground signals.
The first requirement is easy to satisfy and translates into the need for picking sky locations with angular separations greater than the worst (largest) sky bin resolvable by the search, as explained in \sect{resolution}.
As we show below, the second requirement can also be satisfied by picking off-source locations far-enough away from the true position of the source.

\newcommand{\skylocon}{\skyloc_{\rm on}}
\newcommand{\skylocoff}{\skyloc_{\rm off}}

\subsubsection{Signal contribution to sky-shifted statistic}

Let us first examine the conditions under which sky-shifting effectively removes contributions from real continuous waves.
For simplicity, consider a signal of fixed frequency ($\fgw$) originating from some known location ($\skylocon$).
Now imagine heterodyning the data containing it by using a mismatched timing correction corresponding to some shifted sky location ($\skylocoff$), as proposed in \sect{offsourcing}.
In full analogy to \eq{data}, we would then obtain binned data like:
\beq
B (t) = \Lambda(t) e^{2\pi i \fgw \Delta \tau (t)} + n(t) \equiv \Lambda'(t) + n(t)\, , 
\eeq
where we no longer assume the instrumental noise $n(t)$ is normally distributed, and where
\beq \label{eq:deltat}
\Delta \tau (t; \deltasky) \equiv \tau(t; \skylocon) - \tau (t; \skylocoff)
\eeq
represents the timing-correction mismatch between the two sky locations with angular separation $\deltasky = \skylocoff - \skylocon$ [cf. \eq{time}].
As a proxy for a generic search statistic, consider the evaluation of a simple inner product between the data and the expected template:
\beq
\left\langle B(t) \mid \Lambda(t) \right\rangle = \left\langle n(t) \mid  \Lambda(t)\right\rangle + \left\langle \Lambda'(t) \mid  \Lambda(t)\right\rangle .
\label{eq:sum}
\eeq
Our goal is to estimate the distribution of the overlap between the noise and the template, $\left\langle n(t) \mid  \Lambda(t)\right\rangle$, by studying our proxy statistic, $\left\langle B(t) \mid \Lambda(t) \right\rangle$.
Consequently, we would like the contribution of the true signal, $\left\langle \Lambda'(t) \mid  \Lambda(t)\right\rangle$, to \eq{sum} to be sufficiently small to be effectively undetectable.

Explicitly, the contribution of the signal to the inner product of \eq{sum} can be written in terms of a time integral over the observation time $T$,
\beq
\left\langle \Lambda'(t) \mid  \Lambda(t)\right\rangle = \left| \int_0^T \Lambda^2(t)\, e^{2\pi i \fgw \Delta \tau (t; \deltasky)} \infd t \right| \, ,
\label{eq:int1}
\eeq
where $\Lambda^2(t) \equiv \Lambda^*(t) \Lambda(t)$.
The first key feature of this result is that a signal with greater signal-to-noise ratio (SNR) will tend to contaminate the sky-shifted statistic more strongly.
This is not at all surprising: a strong signal can be detected even if there is a small error in its assumed sky location, because enough coherent power can remain even after the timing correction spreads it over several frequency bins.
In fact, \eq{int1} is bounded from above by the squared-norm of the signal template,
\beq \label{eq:ipbound}
\left\langle \Lambda'(t) \mid  \Lambda(t)\right\rangle \leq \left\langle \Lambda(t) \mid  \Lambda(t)\right\rangle = \left| \int_0^T \Lambda^2(t)\, \infd t \right| ,
\eeq
which, for a flat power spectrum, is directly proportional to the square of the SNR.

The second relevant feature of \eq{int1} is that the dependence of $\left\langle \Lambda'(t) \mid  \Lambda(t)\right\rangle$ on sky location will be determined solely by the angular structure of $\Delta\tau$, and how well that can be resolved given $\fgw$ and $T$ (see \sect{resolution}).
The inequality of \eq{ipbound} is, of course, saturated if and only if the shifted location is such that $\Delta\tau(t;\deltasky)=0$ at all times:
this takes place, for instance, if the ``shifted'' location is really just the original location of the source ($\deltasky=0$).
On the other hand, for most other values of $\deltasky$ and for the range of $\fgw$'s we are interested in, the exponential term in \eq{int1} is highly oscillatory---this means that we should expect $\left\langle \Lambda'(t) \mid  \Lambda(t)\right\rangle$ to quickly vanish as we move away from the true location of the source.
This is consistent with the sky-bin definition given in \sect{resolution}, from which it is possible to see that the sky-bin size decreases with increasing frequency.
In other words \eq{int1} is representative of the templates density lattice for a GW search \cite{2013PhRvD..88l3005W, 2015PhRvD..92h2003W, 2018PhRvD..98j2003M}.

\subsubsection*{Angular pattern and magnitude}

\begin{figure}
\includegraphics[scale=0.42]{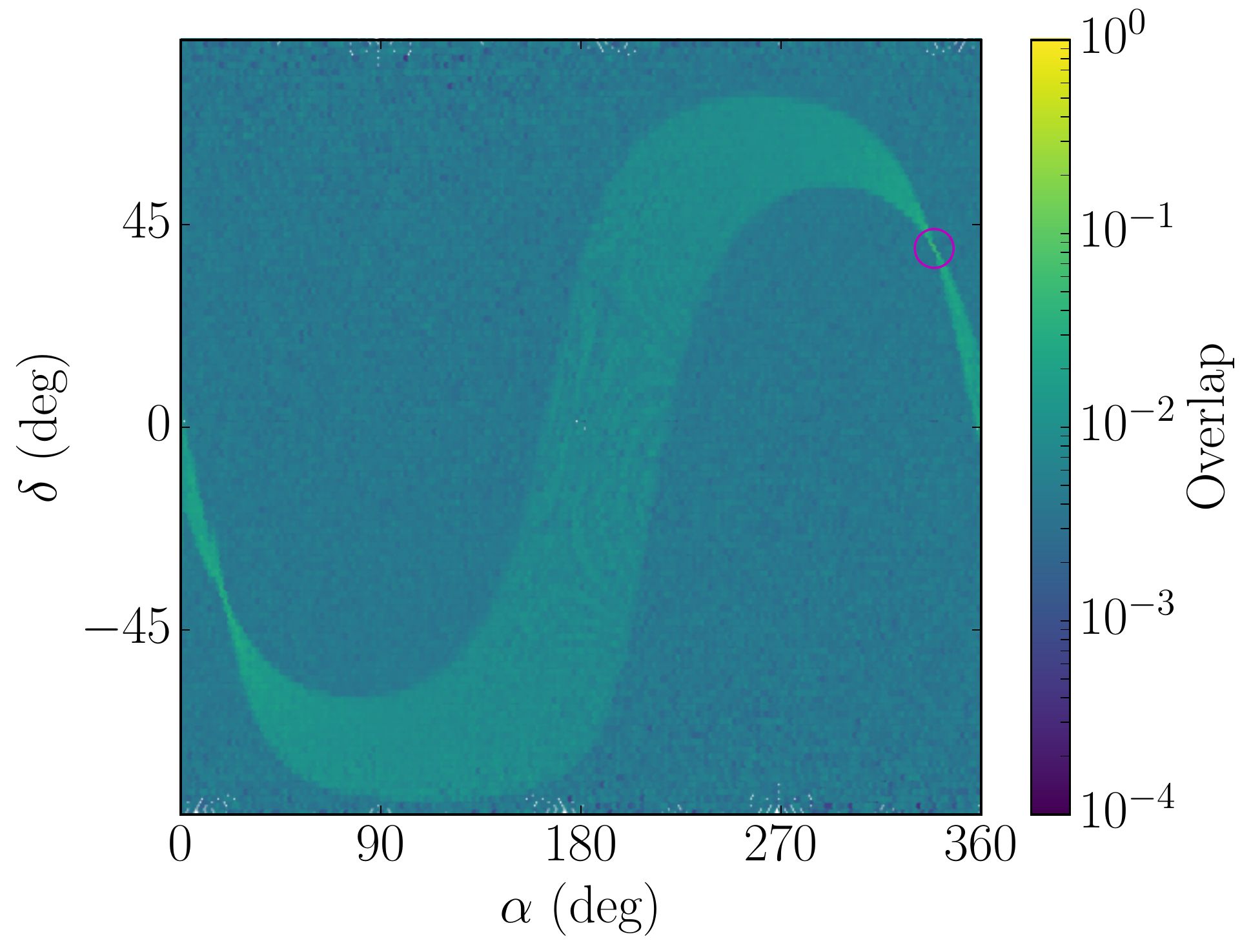}
\caption{Value of the reheterodyned correlation overlap of \eq{overlap} over the entire sky, as a function of right ascension ($\alpha$) and declination ($\delta$). The data contain a simulated signal with $\mathrm{SNR} = 70$ located at $\alpha=22^{\rm h}35^{\rm m}40.73^{\rm s}$, $\delta=39\degree40'44.76''$ (magenta circle).
}
\label{fig:rom_plot}
\end{figure}

The detailed angular structure of our proxy sky-shifted statistic is represented in \fig{rom_plot}.
To produce this plot, we began with a set of binned data, \eq{data}, containing {\em Gaussian} noise and a very strong ($\text{SNR}=70$) simulated signal from an arbitrary sky location on the ecliptic plane (indicated by a magenta circle).
We then reheterodyned these data assuming different (sky-shifted) locations covering the whole sky, and for each instantiation computed the overlap (normalized cross-correlation),
\beq \label{eq:overlap}
\text{Overlap} = \frac{\left\langle B'(t) \mid B(t)\right\rangle}{\left\langle B(t) \mid B(t)\right\rangle}
\approx\frac{\left\langle \Lambda'(t) \mid \Lambda(t)\right\rangle}{\left\langle \Lambda(t) \mid \Lambda(t)\right\rangle}  , 
\eeq
between the sky-shifted data, $B'(t)$, and the on-source data, $B(t)$.
This quantity (shown in color in \fig{rom_plot}) represents the normalized contribution of the injected signal to the sky-shifted statistic for different sky locations, as desired.
This is because $\left\langle n'(t)\mid n(t)\right\rangle\approx 0$ for Gaussian noise, yielding the approximate equality in \eq{overlap}.

\begin{figure*}
\includegraphics[width=2\columnwidth]{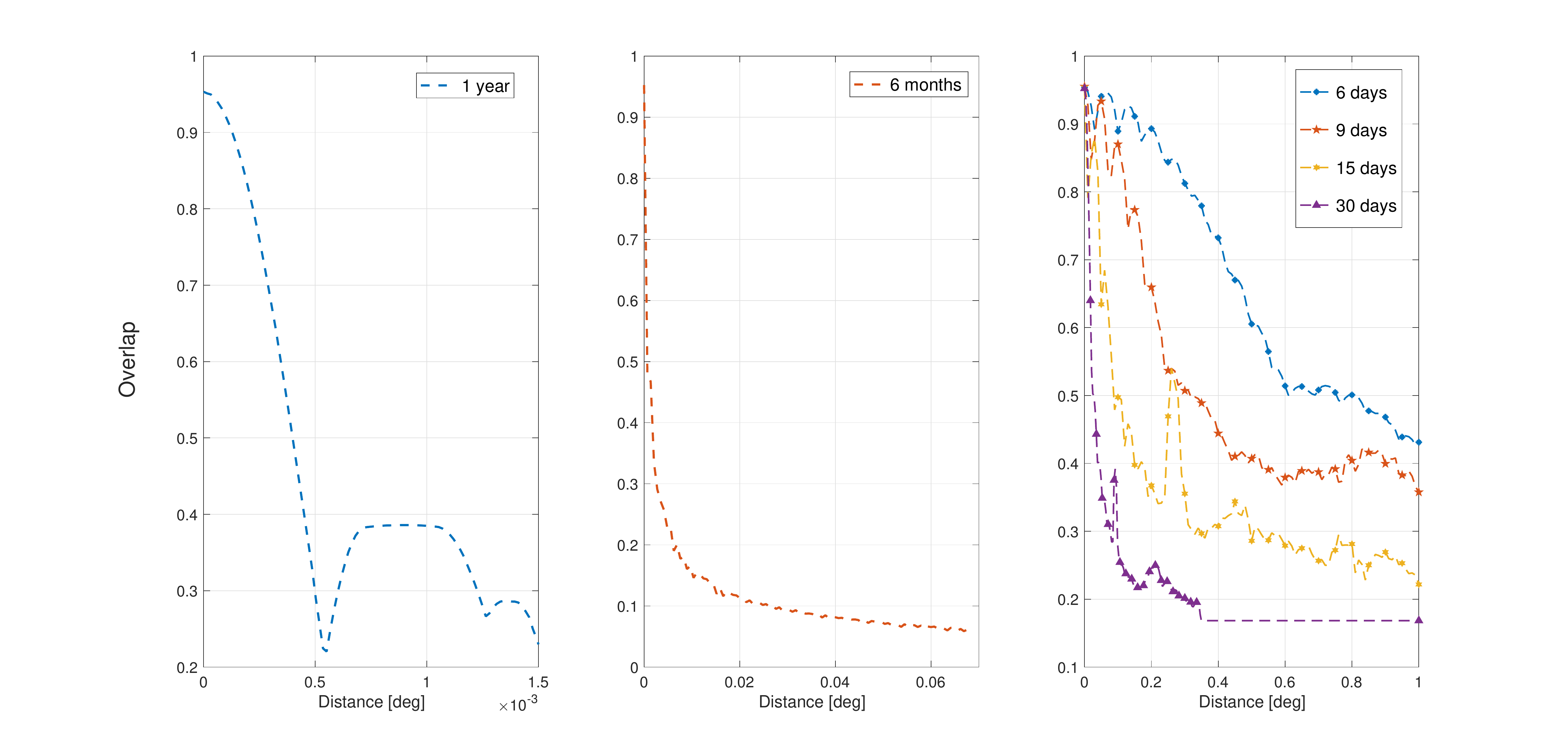}
\caption{
{\it Left panel:} Overlap function, \eq{overlap}, computed for a software injection with the same parameters as Fig.~\ref{fig:rom_plot} for 1 year of integration and close to the source;
the overlap drops very rapidly over an angular distance of ${\sim}5\times 10^{-4}$ deg from the source.
{\it Central panel:} Overlap function computed for a software injection with the same parameters as Fig.~\ref{fig:rom_plot} for 6 months of integration;
the overlap drops over a larger distance with respect to 1 year of integration.
{\it Right panel:} Overlap function computed for a software injection with the same parameters as Fig.~\ref{fig:rom_plot} for various integration times (see legend);
Shorter integration times induce correlations over larger angular scales.
The on-source overlap (null distance) is not identically $1$ due to the presence of noise.
}
\label{fig:corr_dist}
\end{figure*}

As expected, the contribution of the signal falls off steeply as we move away from the source location: while the normalized overlap of \fig{rom_plot} equals unity if $\deltasky=0$ (center of the magenta circle), it is orders of magnitude smaller for all other choices of $\skylocoff$.
The rest of the structure in this plot reflects the symmetries of the timing correction \eq{time}, which are themselves dominated by the \romer{} term in \eq{romer}: locations across lines of fixed ecliptic latitude remain somewhat correlated to the on-source location, and the whole pattern is symmetric under reflections through the ecliptic (see \sect{resolution} for more details).
This important observation means that, for any given on-source location, we will want to sample our sky-shifted points from only one of the ecliptic hemispheres.
This may be achieved by laying out a grid in the sky, or by picking sky locations randomly.
In either case, one must ensure uniqueness of the chosen point by enforcing a minimal separation set by the overlap function in Fig. \ref{fig:corr_dist}.

For a sufficiently loud signal, sky bins neighboring the source will yield contaminated sky-shifted data (i.e.~data that still contains measurable coherent power due to the on-source signal).
Unlike the angular dependence, the overall magnitude of this contamination will be determined by the SNR of the signal and, as such, will depend on the integration time and intrinsic amplitude.
For the same example as in \fig{rom_plot}, \fig{corr_dist} shows the rate at which the overlap with the on-source location decreases as one moves at constant ecliptic latitude away from the source and for different integration times.

Because latitude is held constant in this plot, \fig{corr_dist} represents the slowest-possible decrease in the contamination by this source (cf.~\fig{rom_plot}).
Furthermore, this example was chosen to have very high SNR and to lie on the ecliptic plane, where the sky resolution is poorest (cf.~\sect{resolution})---all of which makes this close to a worst-case scenario.
In spite of this, the overlap vanishes quite quickly, plateauing far away from the source at a value of the order $0.01$.
For realistic (low SNR) CW signals a $1\%$ overlap is small enough to make the signal term in \eq{sum} negligible with respect to the noise.
Hence for  1 month of data integration or more, setting sky-shifting separation to 1 deg is enough to remove any measurable correlation between sky bins in any realistic situation.
Examples of this are given in \sect{results} below.

\subsubsection{Contaminated backgrounds} \label{sec:contamination}

An analysis that draws part of the background from a measurably contaminated region may underestimate the significance of the true signal, but never overestimate it.
This is because a contaminated background will show artificial tails towards higher values of the detection statistic, due to coherent power left over after sky-shifting in some of the ``noise-only'' instantiations.
Thus, in a sense, such an analysis would, at worst, be conservative.
Yet, as we will show in \sect{results}, a signal that is sufficiently loud to cause such contamination over a non-negligible region of the sky will itself yield an on-source detection statistic that is significantly higher than any of the contaminated-background tails.
Therefore, the significance (e.g.~$p$-value) assigned to such a signal will be the same with or without the tails.

In any case, sky bins in the immediate vicinity of the source may always be removed from the background estimation to prevent contamination.
However, the excision of a large area of the sky will have the detrimental consequence of effectively reducing the number of sky bins available for background estimation.
Furthermore, such procedure is only justified if we (implicitly) assume that the on-source data do contain an astrophysical signal.
In a way, this is analogous to how a very loud CBC signal may pollute the time-slid background in searches for transient gravitational waves (e.g.~see caption to Fig.~3 in \cite{o1bbh}, or Fig.~7 in \cite{TheLIGOScientific:2016qqj}).
In that case, the standard procedure has been to first compute significances with the ``polluted'' background to determine whether the zero-lag detection candidate is a real signal, and only if that is the case remove it from the background.%
\footnote{These two kinds of background are known colloquially as with and without ``little dogs'', since this distinction first arose during the analysis of an injection in the direction of \emph{Canis Major} \cite{Colaboration:2011np}.}
The same can be done here if necessary.

In summary, we conclude that sky-shifting, as described in \sect{offsourcing}, is a viable method for estimating the background distribution of detection statistics in targeted searches for continuous waves, as long as the shifted sky-locations are chosen such that: ({\em i}) they are distributed over only half the sky; ({\em ii}) the angular distance between them is no shorter than the sky-resolution of the search (cf.~\sect{resolution}).
This will guarantee that the different draws from the background distribution (obtained from different shifted sky locations) are distinct and uncontaminated by a true signal, if present.

\section{Analysis} \label{sec:results}

\begin{table*}
\caption{Parameters for the case-study signals (\sect{results_cases}).}
\label{tab:pulsars}
\begin{ruledtabular}
\begin{tabular}{ l c c c c l l}
 & $f_{\rm GW}$ {\scriptsize (Hz)} & $\alpha$ & $\delta$ & Figs. & Data & Comment \\
\midrule
J0534+2200 & 59.33 & $5^{\rm h}34^{\rm m}31.97^{\rm s}$ & $22\degree 00'52.07''$ & \ref{fig:crab_medium}, \ref{fig:crab_strong} & Gauss.~design {\scriptsize (H,L,V)} & Assumed $\iota = 61.3\degree$, $\psi=124.0\degree$  \\
J1932+17 & 47.81 & $19^{\rm h}32^{\rm m}07.17^{\rm s}$ & $17\degree 56'18.70''$ & \ref{fig:j1932+17} & Real O1 {\scriptsize (H,L)}  & Published in \cite{o1cw}  \\
P03 & 108.86 & $11^{\rm h}53^{\rm m}29.42^{\rm s}$ & $-33\degree26'11.77''$ & \ref{fig:pulsar3} & Real O1 {\scriptsize (H,L)}  & Hardware injection \cite{Biwer:2016oyg}  \\
\end{tabular}
\end{ruledtabular}
\end{table*}

We study the efficacy of sky-shifting (\sect{offsourcing}) as a viable method to empirically estimate the background distribution of detection statistics in targeted searches for continuous gravitational waves from known pulsars.
We do this in the context of both the Bayesian (\sect{searches_bayes}) and frequentist 5-vector (\sect{searches_5vector}) analyses, to demonstrate the generality of the approach.
We discuss specific case-studies in \sect{results_cases}, and systematically compare to different methods by computing false-dismissal and false-alarm rates in \sect{results_comparison}.

The following results make use of both simulated and actual noise from interferometric detectors.
In all cases, we begin with a set of data representing the (synthetic or actual) output of a detector after applying the preprocessing required to target some chosen pulsar (e.g.~filtering and downsampling)---these are the on-source data.
We then proceed as described in \sect{offsourcing} to generate multiple new sets of sky-shifted data, and then evaluate the distribution of the detection statistic over all such instantiations (excluding the original, on-source data).
We can then compare the value of the on-source statistic to the sky-shifted background, as we would in a real analysis.

\subsection{Case studies} \label{sec:results_cases}

\begin{figure}
\includegraphics[width=\columnwidth]{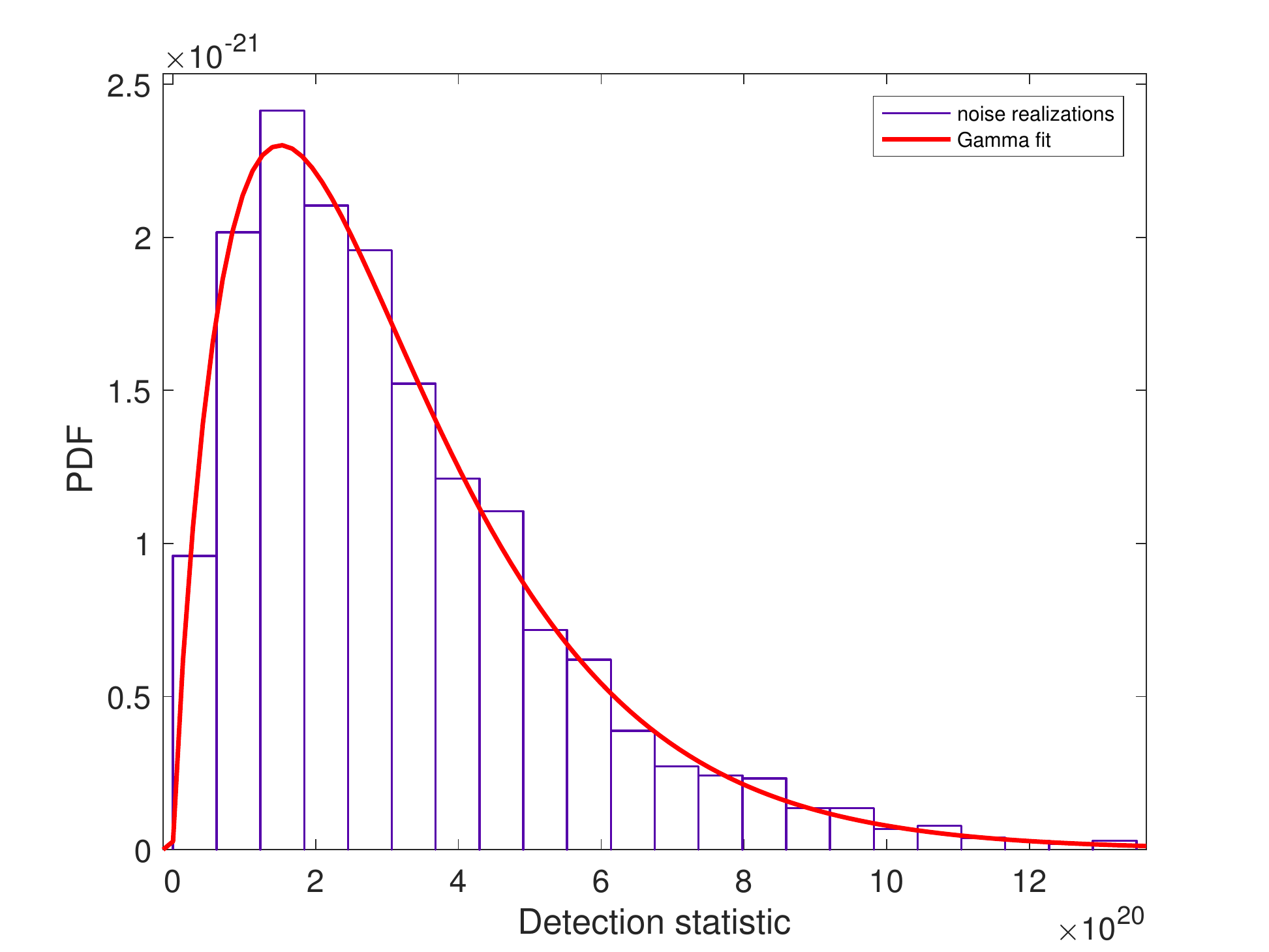}
\caption{Histogram of the noise-only distribution obtained using the sky-shifting method in case of Gaussian noise (no signal). Red line: best fit given by a $\Gamma$ distribution with $a=1.95$.
In this test case the hypothetical source was assumed at a frequency $f\approx 108.85$ Hz with a spin-down  $\dot{f}\approx 10^{-17}$ Hz/s.
The on-source position was chosen on the ecliptic plane ($\alpha=22^{\rm h}35^{\rm m}40.73^{\rm s}$, $\delta=39\degree40'44.76''$) as in \fig{rom_plot}. The search assumed a single detector at the LIGO Hanford site.}
\label{fig:5-vec_Gaussian}
\end{figure}

Here we provide several concrete examples of sky-shifting at work in the presence of pure noise, realistic signals and strong signals, as summarized in Table \ref{tab:pulsars}.
Background distributions are estimated from $10^4$ shifted locations in the hemisphere of the source.
The simulations of Gaussian noise (\sect{results_cases_Gaussian}) were carried out assuming an observation time of 6 months and PSDs corresponding to the aLIGO and Virgo design sensitivities.
With the exception of \fig{5-vec_Gaussian}, the simulated data for LIGO Hanford (``H''), LIGO Livingston (``L'') and Virgo (``V'') detectors were then analyzed coherently with the Bayesian method of \sect{searches_bayes}, to obtain the signal vs noise Bayes factors of \eq{bsn} as our detection statistic.

The examples with real instrumental noise correspond to LIGO's first observation run (O1).
The data streams start on 2015 Sep 11 at 01:25:03 UTC for Hanford and 18:29:03 UTC for Livingston, and finish on 2016 Jan 19 at 17:07:59 UTC at both sites.
The first example consists of data prepared for the pulsar PSR J1932+17, for which search results were presented in \cite{o1cw}.
All analysis settings are the same as in \cite{o1cw}, except for a log-uniform prior in the signal amplitude (same as in \cite{o1cw-ngr}).
The second example is for a hardware injection presented in \cite{Biwer:2016oyg}.
Both these examples are offered merely to demonstrate the performance of sky-shifting under realistic circumstances---we present no new observational results.

\subsubsection{Pure Gaussian noise} \label{sec:results_cases_Gaussian}

We first demonstrate that sky-shifting works as expected in pure Gaussian noise and in the absence of signal.
In this case, the on-source data are just a set of samples from a Gaussian distribution with zero-mean and standard deviation given by the value of the detector PSD at the GW frequency expected from the targeted pulsar.
The sky-shifting process should correspondingly produce multiple instantiations of independent Gaussian noise, a fact that should be reflected in the resulting background distribution of the detection statistic.
This distribution is shown in \fig{5-vec_Gaussian} for an example using the 5-vector statistic of \eq{5ds}.
When computed over Gaussian noise, it can be shown that this statistic must follow a $\Gamma$ distribution with 2 degrees of freedom \citep{Astone2010}.
\fig{5-vec_Gaussian} shows that this is the case, in agreement with our expectation that sky-shifting should produce independent draws from the background distribution. 

\begin{figure*}
\subfloat{\includegraphics[width=1.05\columnwidth]{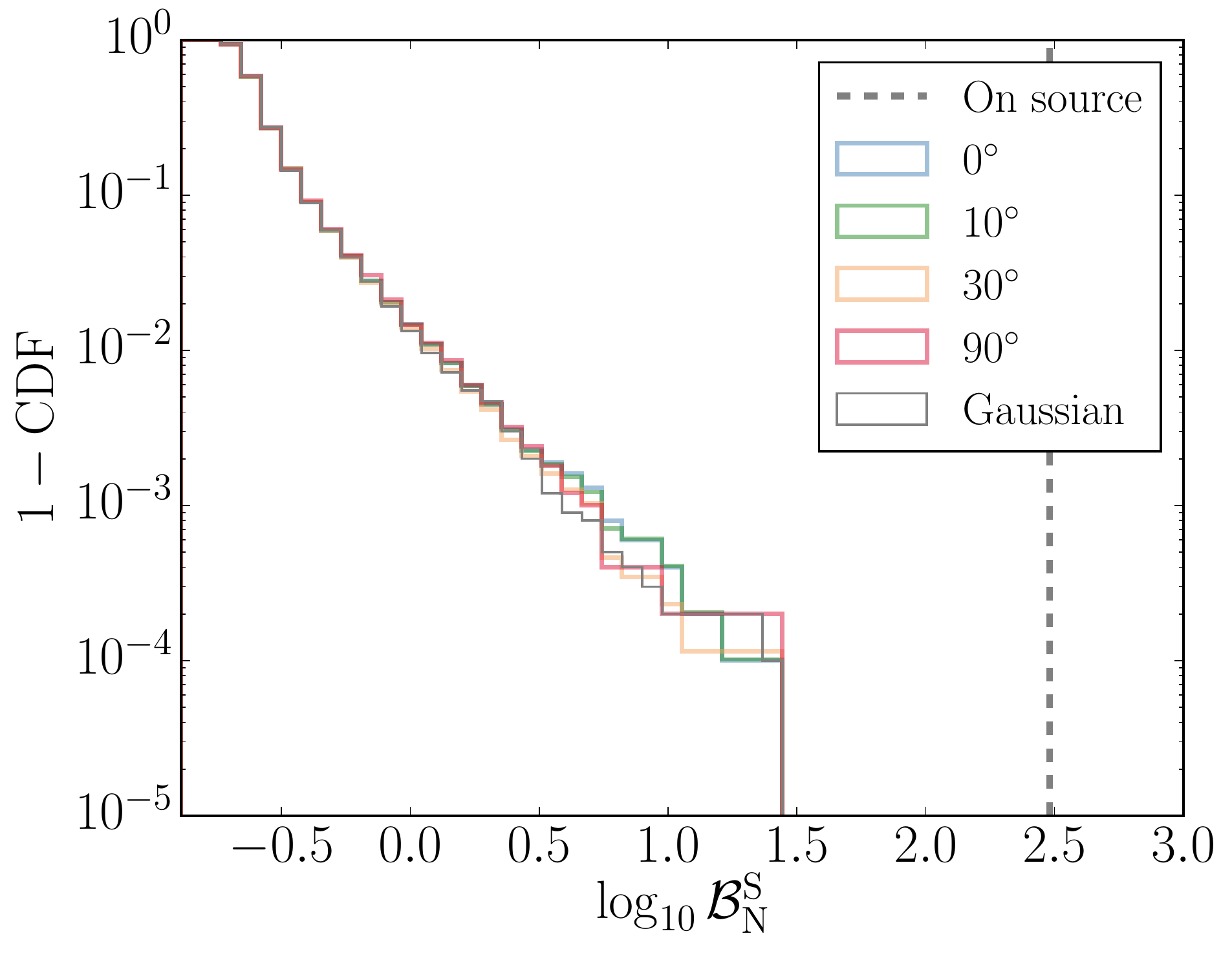}}\hfill
\subfloat{\centering\raisebox{.9cm}{\includegraphics[width=0.92\columnwidth]{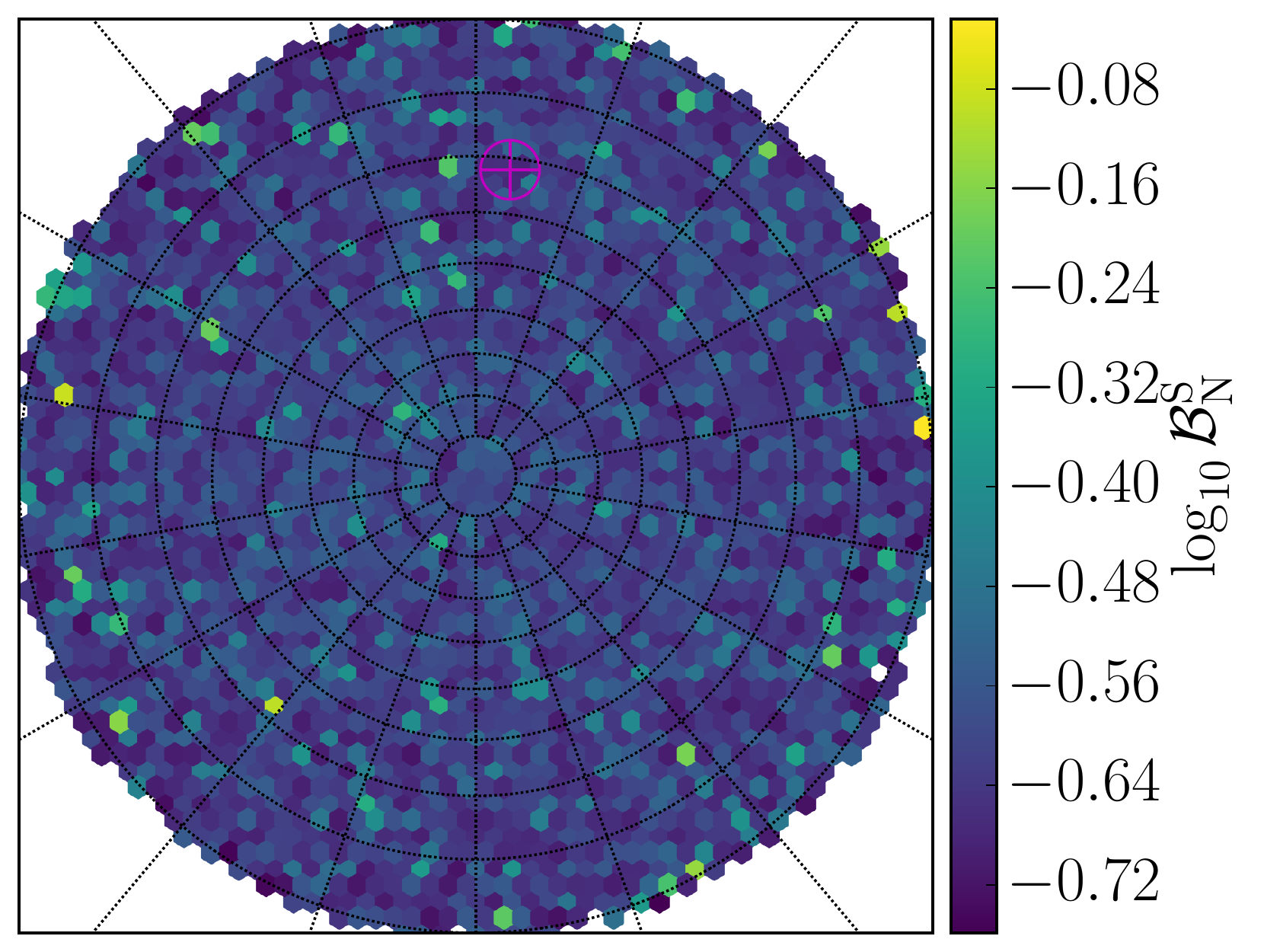}}}
\caption[Realistic Crab signal in Gaussian noise]{{\em Realistic Crab signal in Gaussian noise}.
We simulate a 6-month-long signal with $h_0 = 10^{-26}$ (network SNR 5) and parameters consistent with the Crab pulsar (PSR J0534+2200), inject it in Gaussian noise for aLIGO and Virgo design PSDs, and recover it using the Bayesian analysis of \sect{searches_bayes} (see Table \ref{tab:pulsars}).
The \emph{left panel} shows the survival function ($1-\mathrm{CDF}$) of $\log_{10}\bayessn$, \eq{bsn}, for the sky-shifted background produced from the injected data for different excision areas around the source (different colors, blue and green overlap almost perfectly), as well as from pure Gaussian noise (gray, thin histogram); the on-source statistic for the injection is $\log_{10}\bayessn=2.5$ (thick dashed line), higher than any of the $10^4$ sky-shifted instantiations.
The \emph{right panel} shows the distribution of the sky-shift statistic over the sky in a North-polar stereographic projection, with the Crab's location marked by the magenta crosshairs; the color of each hexagon gives the average of $\log_{10}\bayessn$ over several sky bins.}
\label{fig:crab_medium}
\end{figure*}

\subsubsection{Injections in Gaussian noise}

Ideally, the background distribution should be unaffected by the presence of a signal: while the value of the on-source statistic should rise to reveal it, the sky-shifted values should be insensitive to it.
We demonstrate that this is the case by injecting signals of different amplitudes in Gaussian noise.
We simulate a signal from the Crab pulsar (PSR J0534+2200) as seen by three advanced detectors (H, L, V) at design sensitivity over 6 months, and recover it using the Bayesian method of \sect{searches_bayes}.

We first choose a realistic signal amplitude of $h_0 = 10^{-26}$, which is weak enough to be consistent with the latest upper limits for this source \cite{Abbott:2020lqk}, but strong enough to yield a nonnegligible $\text{network-SNR} = 5$ for the chosen PSDs and observation time.
The data containing the injected signal are then reheterodyned for $10^4$ shifted sky-locations to yield the survival function in \fig{crab_medium}, i.e.~the complement of the cumulative density function (CDF), $1-\mathrm{CDF}$.
Each colored trace in this figure represents the distribution of sky-shift background computed from the Northern celestial hemisphere, {\em excluding} any points closer to the source than the indicated angular distance, i.e.~excluding points with $(\alpha,\, \delta)$ such that $|\alpha-\alpha_\star|<\Delta$ and $|\delta-\delta_\star|<\Delta$ with a ``$\star$'' indicating the true location of the Crab and $\Delta$ one of the values given in the legend of \fig{crab_medium}: $0\degree$ (blue), $10\degree$ (green), $30\degree$ (yellow) or $90\degree$ (red).
In particular, the blue curve corresponds to background from points sampled over the whole hemisphere, while the red curve corresponds to points sampled over the half-hemisphere not containing the source.

In this case, the choice of sky-region does not have a strong effect on the background: we may take advantage of the whole hemisphere, getting quite close to the source (as allowed by the frequency resolution of this search).
In fact, note that the blue and green curves in \fig{crab_medium} are essentially identical.
For reference, the distribution of the sky-shift statistic over the whole Northern sky ($0\degree$ curve on the left) is represented on the right panel of \fig{crab_medium} via a stereographic map, with the true location of the source indicated by magenta crosshairs.

As expected, the background produced via sky-shifting is practically indistinguishable from results in pure Gaussian noise (gray, thin histogram).
Indeed, these two samples yield a Kolmogorov-Smirnov (KS) $p$-value of 0.77, favoring the hypothesis that they were both drawn from a common distribution.
This agreement is in spite of the fact that the on-source statistic (dark gray-dashed line) takes a significantly-increased value, revealing the presence of the injection.
Completely ignoring the intrinsic probabilistic meaning of $\bayessn$, a background like the blue curve in the left panel of \fig{crab_medium} would allow us to place a $p$-value of at most $10^{-4}$ on the null hypothesis that the on-source data are noise.

\begin{figure*}
\subfloat{\includegraphics[width=1.05\columnwidth]{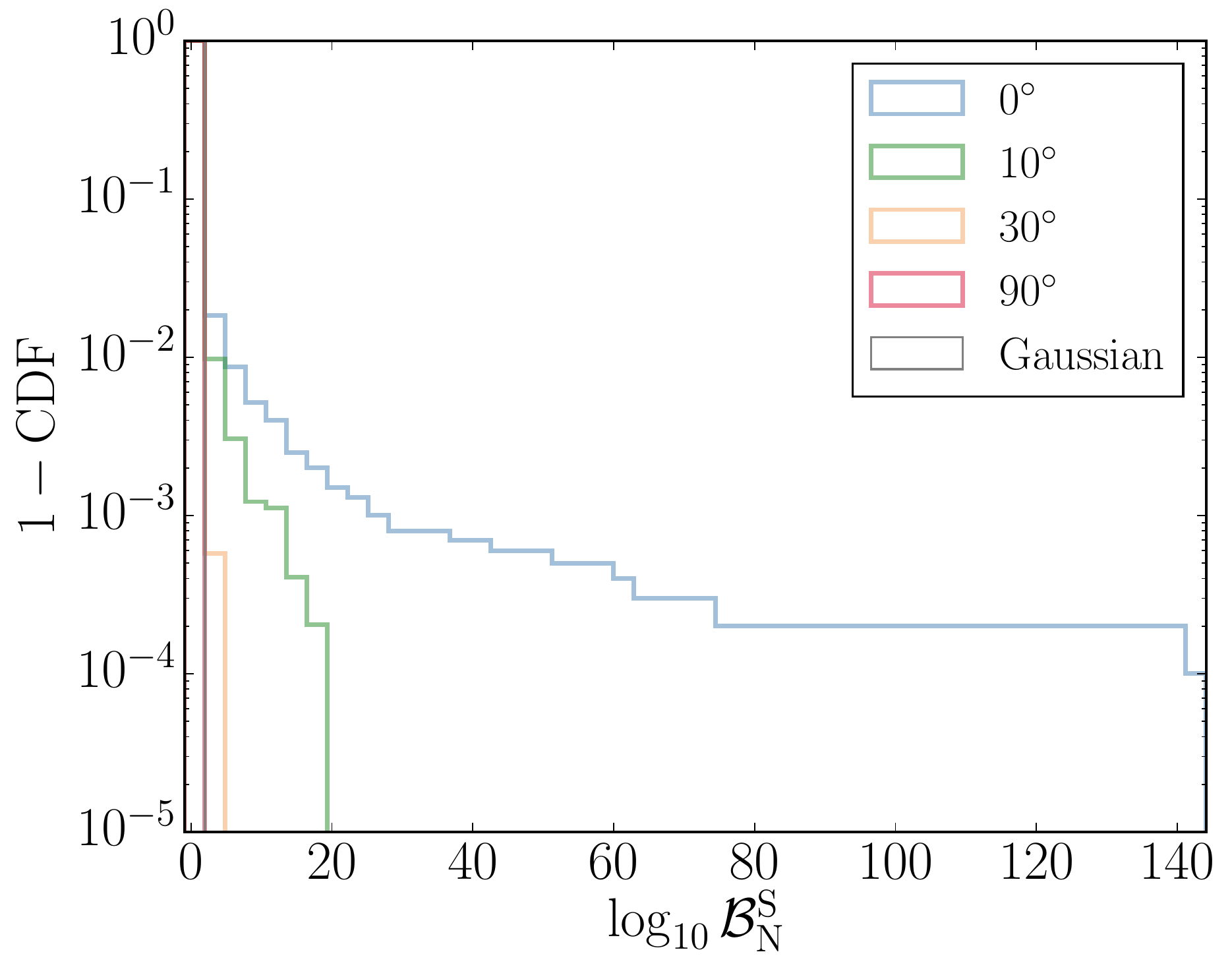}}\hfill
\subfloat{\centering\raisebox{0.9cm}{\includegraphics[width=0.88\columnwidth]{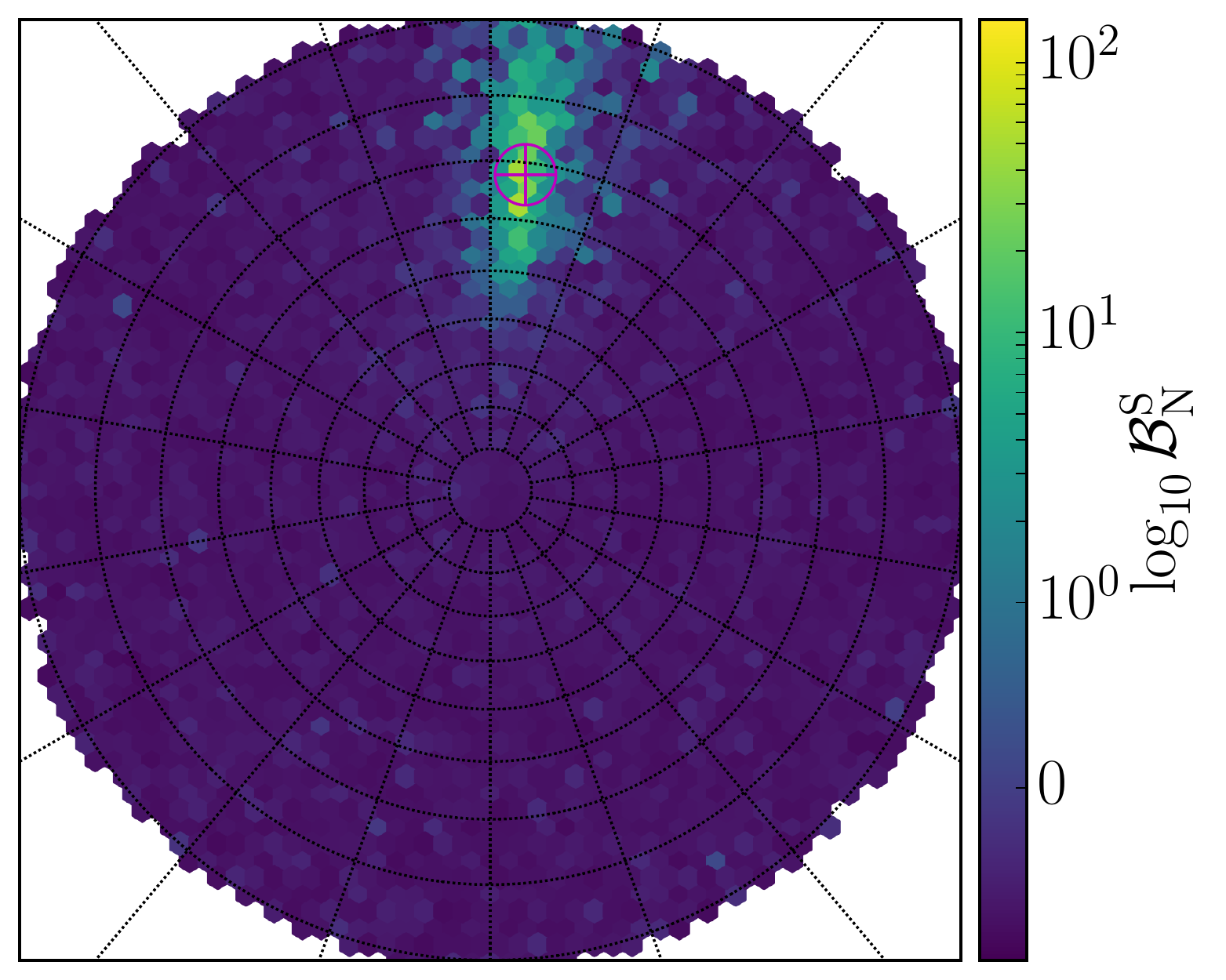}}}
\caption[Very loud Crab signal in Gaussian noise]{{\em Very loud Crab signal in Gaussian noise}.
We simulate a 6-month-long signal with $h_0 = 1.4\times10^{-24}$ (network SNR 700) and parameters consistent with the Crab pulsar (PSR J0534+2200), inject it in Gaussian noise for aLIGO and Virgo design PSDs, and recover it using the Bayesian analysis of \sect{searches_bayes} (see Table \ref{tab:pulsars}).
The \emph{left panel} shows the survival function ($1-\mathrm{CDF}$) of $\log_{10}\bayessn$, \eq{bsn}, for the sky-shifted background produced from the injected data for different excision areas around the source (different colors), as well as from pure Gaussian noise (gray, thin histogram); the on-source statistic for the injection is $\log_{10}\bayessn=9\times10^4$, which is vastly higher than any of the $10^4$ sky-shift instantiations (off the scale).
The \emph{right panel} shows the distribution of the sky-shifted statistic over the sky in a North-polar stereographic projection, and with the Crab's location marked by the magenta crosshairs; the color of each hexagon gives the average of $\log_{10}\bayessn$ over several sky bins, in semi-log scale linearly interpolated between $(-1,1)$.}
\label{fig:crab_strong}
\end{figure*}

As anticipated in \sect{blinding}, there is a limit to how loud the injection can be without noticeably contaminating nearby sky-bins and, therefore, biasing the background distribution obtained through sky-shifting.
However, this threshold is quite high: for the same detector configuration as above, we find that the injection must reach $h_0 \sim {\cal O} \left(10^{-24}\right)$, or a network-SNR ${\sim}700$ at design sensitivity, before sky-shifting is unable to effectively remove it.
We show an example of this in \fig{crab_strong} for a signal from the Crab pulsar with $h_0 =  1.4\times 10^{-24}$, which roughly corresponds to the spin-down limit for this source \cite{o1cw}.%
\footnote{For an isolated pulsar (no accretion), the spin-down limit is the maximum power that could possibly be emitted in gravitational waves given the observed decay in the pulsar's angular momentum.}
This time, as seen from the panel on the left, the full-hemisphere sky-shifting distribution (blue curve) is visibly inconsistent with a pure-Gaussian background (gray, thin curve), and a KS $p$-value of $10^{-7}$ strongly disfavors a shared distribution between the two sample sets.
From the right panel, it is clear that the culprits are noticeably-contaminated sky locations in the neighborhood of the source (magenta crosshairs).
These polluted sky bins are arranged in the same pattern predicted in \fig{rom_plot}, although under a different guise due to the logarithmic color scale.

\begin{figure*}
\subfloat{\includegraphics[width=1.02\columnwidth]{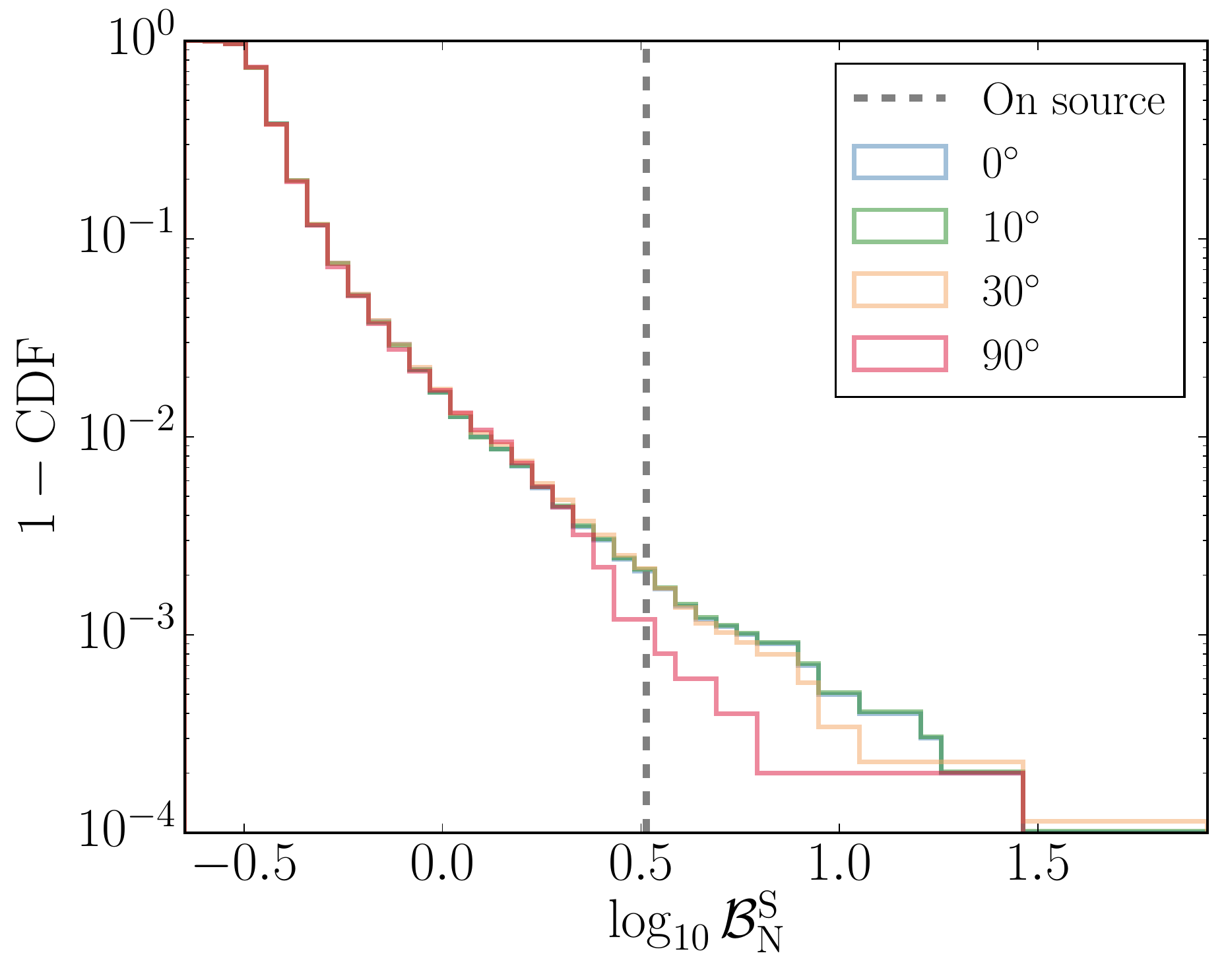}}\hfill
\subfloat{\centering\raisebox{0.9cm}{\includegraphics[width=0.92\columnwidth]{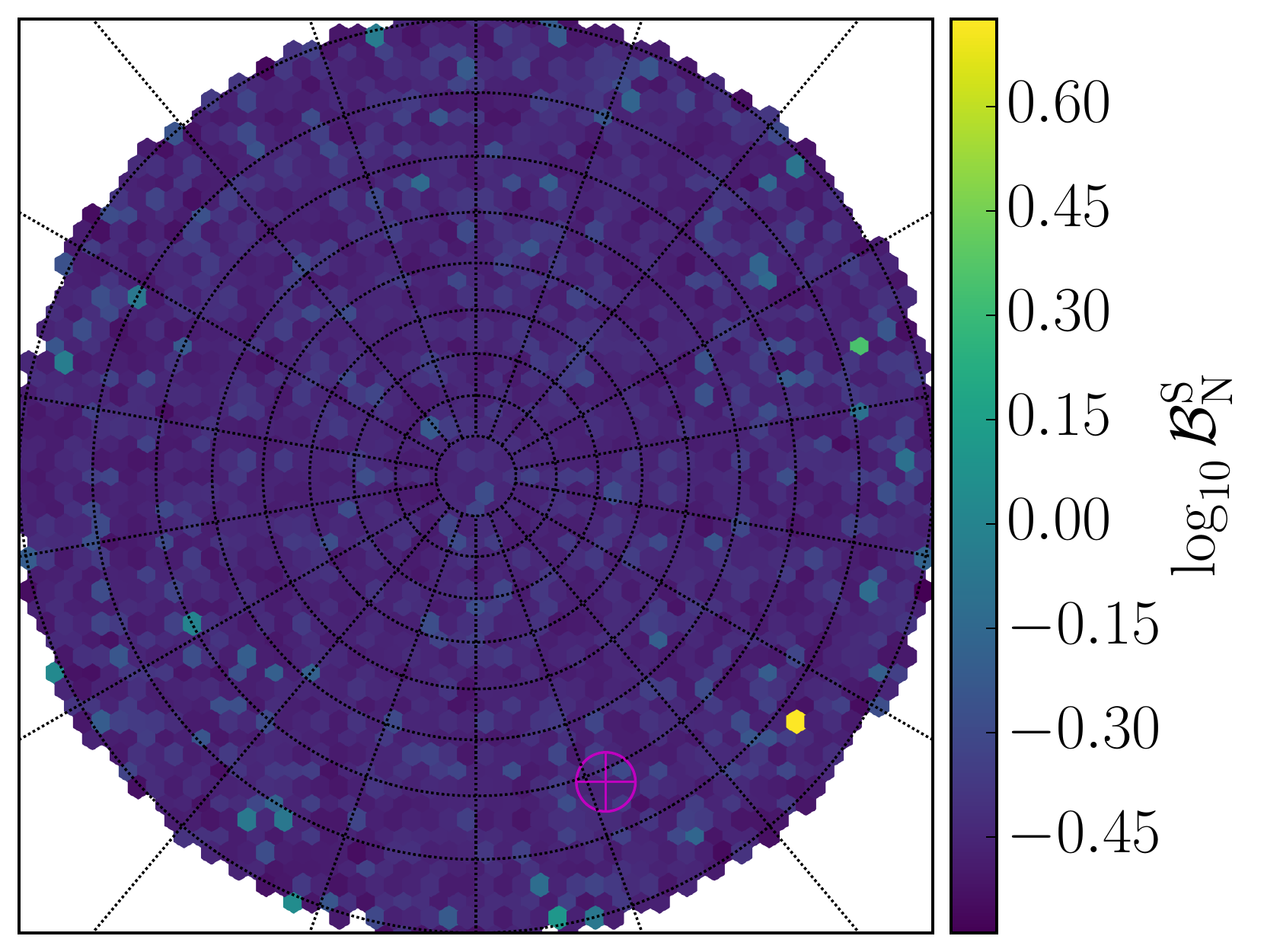}}}
\caption[Real O1 noise for J1932+17]{{\em Real O1 noise for J1932+17.}
Off-sourced background produced from real O1 LIGO data prepared for PSR J1932+17, analyzed coherently with the Bayesian method of \sect{searches_bayes} (see Table \ref{tab:pulsars}).
The \emph{left panel} shows the survival function ($1-\mathrm{CDF}$) of $\log_{10}\bayessn$, \eq{bsn}, for the sky-shifted background for different excision areas around the source (different colors); the on-source statistic for this pulsar is $\log_{10}\bayessn=0.5$ (vertical dashed line), which differs from the result in \cite{o1cw} only due to a log-uniform prior on the signal amplitude.
The \emph{right panel} shows the distribution of the sky-shifted statistic over the sky in a North-polar stereographic projection, with the true location marked by the magenta crosshairs; the color of each hexagon gives the local average of $\log_{10}\bayessn$ over several sky bins.}
\label{fig:j1932+17}
\end{figure*}

\begin{figure*}
\subfloat{\includegraphics[width=1.05\columnwidth]{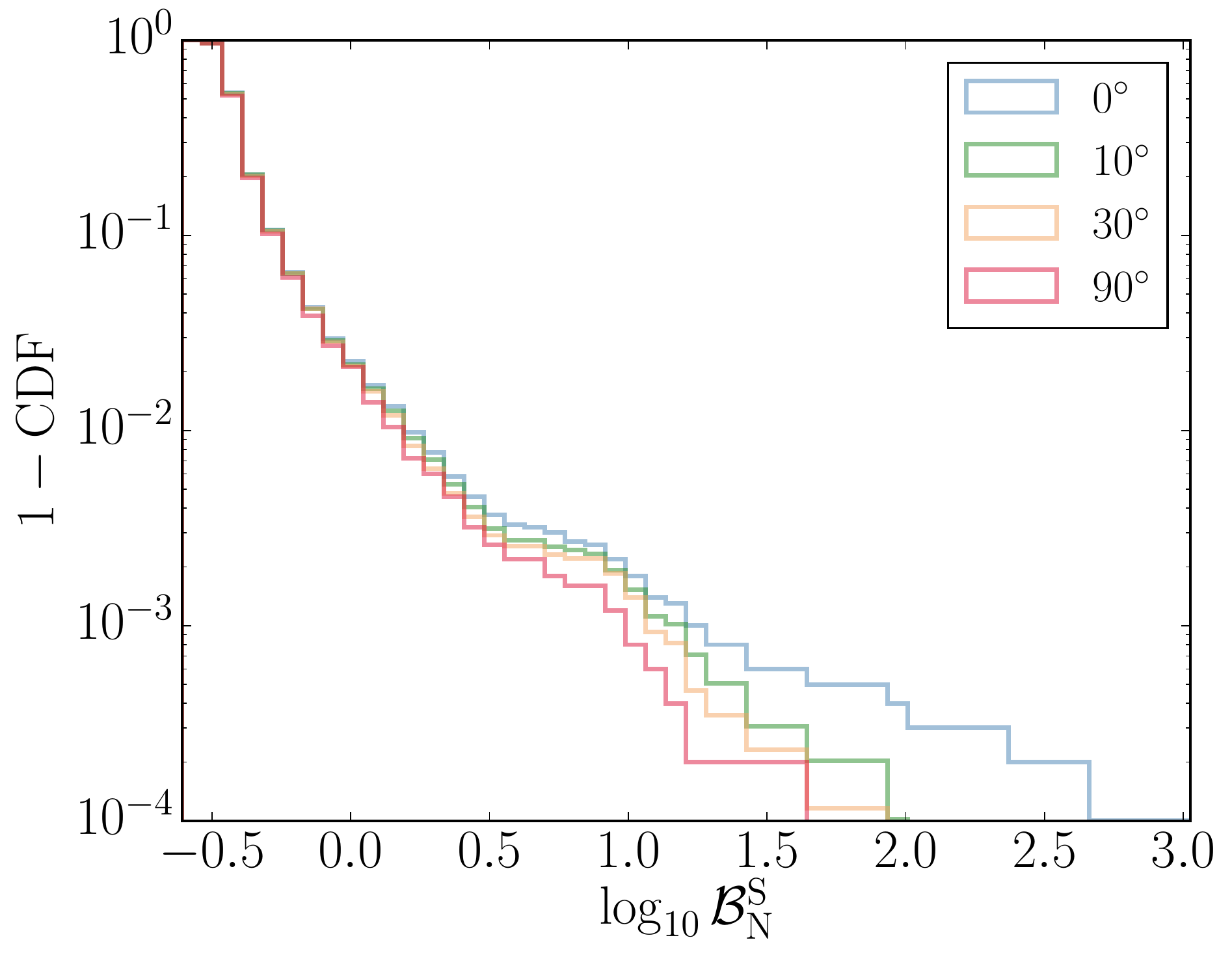}}\hfill
\subfloat{\centering\raisebox{0.9cm}{\includegraphics[width=0.925\columnwidth]{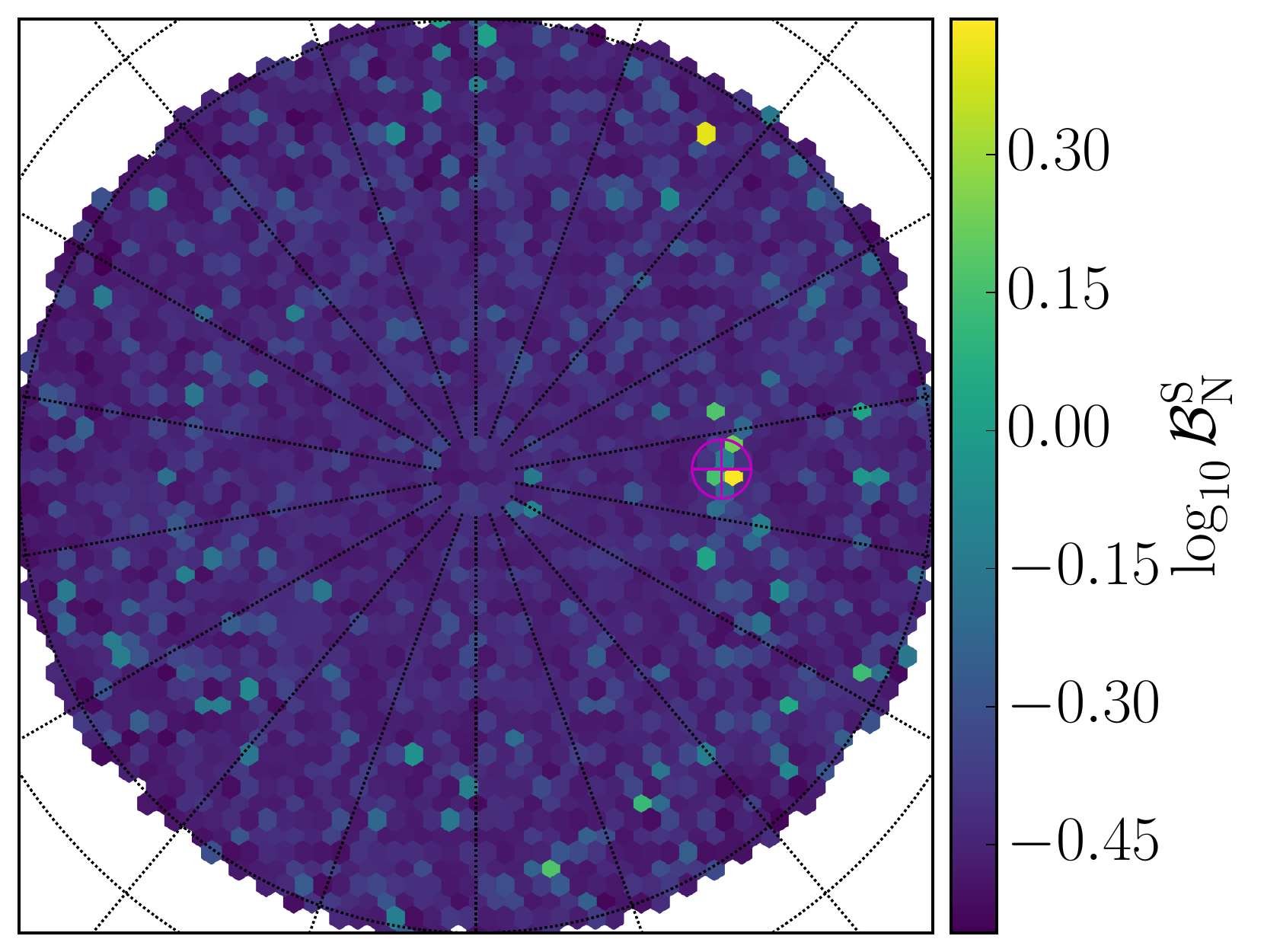}}}
\caption[Real O1 noise for a loud hardware injection]{{\em Real O1 noise for a loud hardware injection (P03).}
Off-sourced background produced from real O1 LIGO data prepared for hardware injection P03 \cite{Biwer:2016oyg}, analyzed coherently with the Bayesian method of \sect{searches_bayes} (see Table \ref{tab:pulsars}).
The \emph{left panel} shows the survival function ($1-\mathrm{CDF}$) of $\log_{10}\bayessn$, \eq{bsn}, for the sky-shifted background for different excision areas around the source (different colors); the on-source statistic for this pulsar is $\log_{10}\bayessn=504$ (off-scale).
The \emph{right panel} shows the distribution of the sky-shift statistic over the sky in a South-polar stereographic projection, with the true location marked by the magenta crosshairs; the color of each hexagon gives the local average of $\log_{10}\bayessn$ over several sky bins.}
\label{fig:pulsar3}
\end{figure*}

As discussed in \sect{blinding}, background contamination can at worst cause us to \emph{underestimate}, never overestimate, the significance of a detection.
Yet, the background is not underestimated in the example of \fig{crab_strong} because the signal is too loud ($\log_{10} \bayessn = 9 \times 10^4$, off the scale of the histogram in \fig{crab_strong}).
This is a general feature: in Gaussian noise, if a signal is loud enough to contaminate a large region of the sky, it will also be louder than the loudest background produced from it.

\subsubsection{Real noise}

The above behavior is replicated in the presence of actual noise from LIGO and Virgo, with the difference that the background naturally shows tails due to the non-Gaussianities in the data.
An example of this is shown in \fig{j1932+17}, which was produced using actual data from aLIGO's first observation run, prepared for the pulsar PSR J1932+17 and with both detectors analyzed coherently using the Bayesian method of \sect{searches_bayes}.
As before, the left panel show the sky-shifted background distributions for different excision areas around the source (different colors).
Note that the excision process does not have any significant impact on the distribution, which is what one would expect in the absence of a very loud signal at the on-source location.
The presence of artifacts in the data becomes apparent in the slower drop of the survival function with respect to, e.g., \fig{crab_medium}.
The on-source value of the signal vs noise Bayes factor for this source was published in \cite{o1cw}, and is marked here by a vertical dashed line---clearly, there is no evidence for a signal in the data.

To study the effectiveness of sky-shifting in detecting a signal in real noise, we analyze data for the hardware injection referred to as ``P03'' in \cite{Biwer:2016oyg}.
Hardware injections are produced by physically actuating on the test masses to mimic the effect of a true gravitational wave, providing a valuable end-to-end test of the instrumental calibration and analysis pipelines. 
In the case of P03, the signal was injected at 108.86 Hz with an amplitude of {$h_0 = 8.2 \times 10^{-25}$} ($\text{network-SNR} = 50$).
As shown on the left panel of \fig{pulsar3}, this signal seems to be sufficiently strong to slightly contaminate the sky bins in its immediate vicinity, but this pollution is easily removed via a narrow excision (compare the blue trace to the rest in \fig{pulsar3}).
In any case, the value of the on-source Bayes factor for this signal is $\log_{10}\bayessn = 504$, which is significantly louder than the loudest background.
Given that $10^4$ sky-shifted noise instantiations were used to estimate the background, this implies that \fig{pulsar3} would allow us to claim a detection of P03 with $p \leq 10^{-4}$ (ignoring the intrinsic probabilistic meaning of $\bayessn$).

\subsection{Comparison to other methods} \label{sec:results_comparison}

\begin{figure*}
\hspace{-0.5cm}
\makebox[\textwidth][c]{\includegraphics[width=1.1\textwidth]{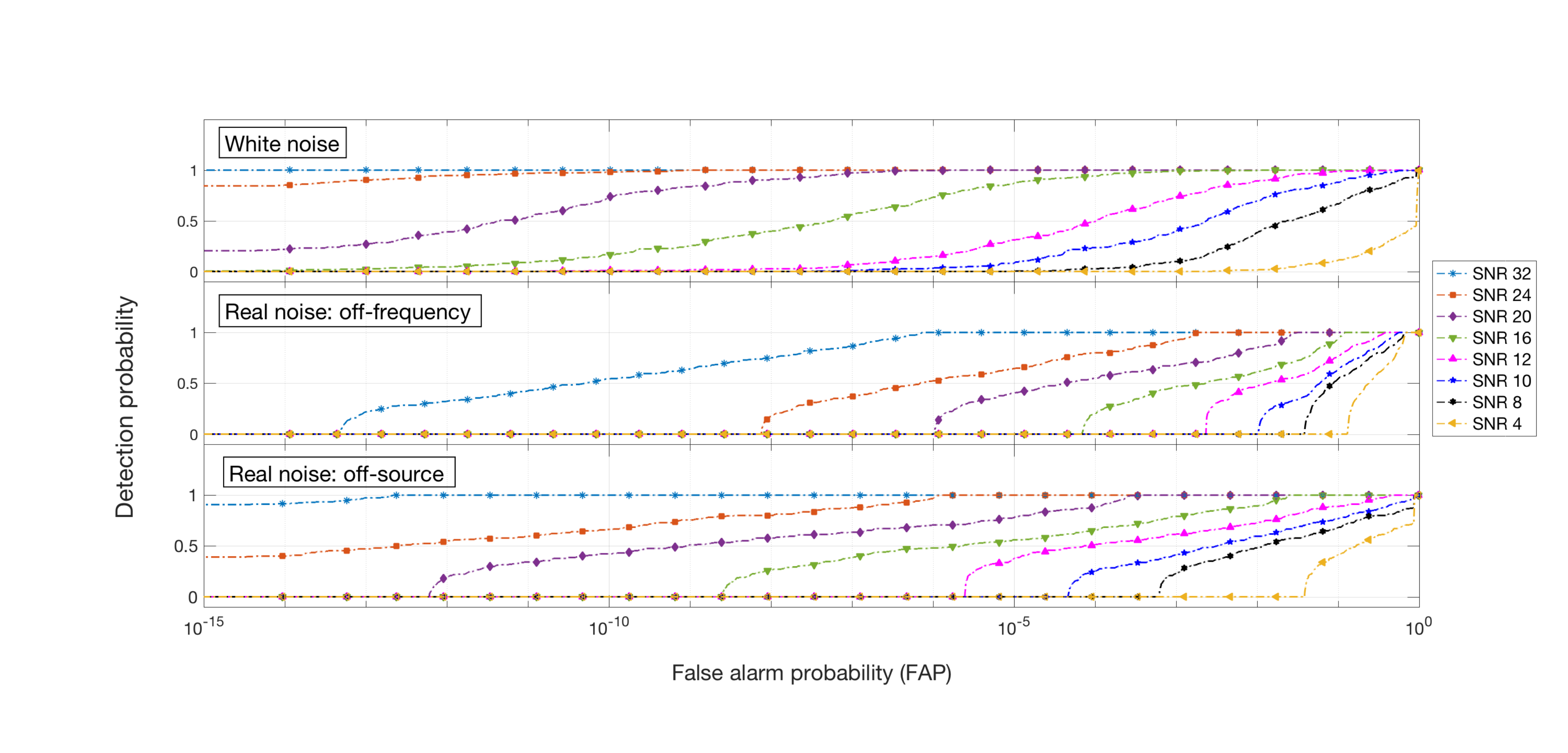}}
\vspace*{-1.0cm}
\caption{\emph{Top panel}: Detection probability vs FAP threshold in Gaussian noise. \emph{Middle panel}: Same but done in non Gaussian noise using the off- frequency method. \emph{Bottom}: Same again but this time using the sky-shift method.}
\label{fig:RR}
\end{figure*}

In order to determine whether sky-shifting offers an improvement over other strategies, we must go beyond specific examples and study false-alarm and false-dismissal rates.
That is, respectively, how likely is sky-shifting to conclude that a noise artifact is a signal (false alarm), versus how likely is it to conclude that a signal is a noise artifact (false dismissal), as a function of confidence level.
We estimate those rates from a large number injections in simulated and actual noise, and use them to directly compare with the standard background-estimation method for the 5-vector search (\sect{searches_5vector}).
In our simulations, we find that sky-shifting can outperform the usual methods in real LIGO data.

\subsubsection{False-dismissal rate}

First, in order to study the false dismissal rate, for a selection of SNRs, we simulate 250 signals  over the sky, with extrinsic parameters $(\psi,\, \eta,\, \phi_0)$ picked randomly over their allowed ranges. 
We then inject these in both idealized (Gaussian) and actual O1 noise for the LIGO Hanford and Livingston detectors (4 months observation time).
In the case of Gaussian noise, we pick the injection frequency and spin-down to be equal to those of P03, a choice that will only affect the specific size of the sky-patches that required for sky-shifting per \eq{skybin}. 
When using real detector data, we set the frequency of the injections to be $54.5$ Hz with no spin-down.
This is because O1 data are known to be polluted by noise artifacts in this frequency band, especially in the Livingston detector \cite{2017PhRvD..96l2006A}, making this a good frequency region to test the performance sky-shifting under realistic circumstances.

In each case, we use the method of \sect{searches_5vector} to obtain the on-source value of the detection statistic, as well as $2\times10^4$ sky-shifted background values with a minimal separation of $0.01 $ deg.
We then compute the number of detected signals as a function of false-alarm probability, i.e.~the number of injections recovered with a detection statistic that is higher than or equal to the value corresponding to certain $p$-value, as established from the empirically-estimated background.
For comparison, we repeat the above procedure but with a background generated via the standard ``off-frequency'' method (mentioned in \sect{searches_5vector} and described in detail in \cite{2017PhRvD..96l2006A}), instead of sky-shifting.
In both cases, the tails of the background distributions are extrapolated for very large values of the detection statistic using an exponential-decay fit.

The results of this study are summarized in the receiver operating characteristic (ROC) curves of \fig{RR}.
Curves in that figure represent the normalized detection rate (detection probability) as a function of $p$-value (FAP) for different choices of injected SNR (different colors and traces) and different methods used to estimate the FAP.
The top panel shows the results from applying the sky-shifting method to the case of Gaussian noise, the middle panel from applying the off-frequency method to real detector data, and the bottom panel from applying the sky-shifting to the same real data as in the middle plot.
As expected, optimal ROCs are obtained in the case of Gaussian noise, for which both methods are equivalent; on real detector data, however, sky-shifting shows better ROCs.

Non-Gaussian artifacts in real noise produce tails in the background distribution, hurting our chances to detect actual signals.
This can be seen by comparing \fig{crab_medium} to \fig{pulsar3}.
This shows in \fig{RR} which indicates that, with those settings, we are only 50\% likely to detect an $\text{SNR}=16$ signal (green curve, down-pointing triangles) with $\text{FAP}=10^{-6}$ (${\sim}5\sigma$) in real O1 noise, but we are 75\% likely to detect it with the same confidence in Gaussian noise.

Nevertheless, \fig{RR} also shows that sky-shifting can outperform traditional methods in the presence of actual instrumental noise, as can be seen by comparing the center and bottom panels.
For instance, the off-frequency method (middle panel) has essentially 0\% chance of detecting an $\text{SNR} = 16$ signal at $\text{FAP}=10^{-6}$, which is dramatically less than the 75\% chance of detecting it via sky-shifting (bottom panel).
In fact, for these settings, sky-shifting is consistently superior at all SNRs.

\subsubsection{False-alarm rate}

Besides assigning high significance to real signals, sky-shifting should also be able to robustly reject outliers arising from non-Gaussianities in the data.
In other words, it should have a low false-alarm rate at any given level of confidence, rejecting artifacts with high probability.
The study of noise features is a very general problem due to the wide morphological range of non-Gaussianities that can be found in the data, which is the same reason why modeling noise likelihoods from first principles is impossible in the first place.
This makes the benchmarking of noise-rejection probability a challenging problem.

\begin{figure*}
\centering
\includegraphics[width=\textwidth]{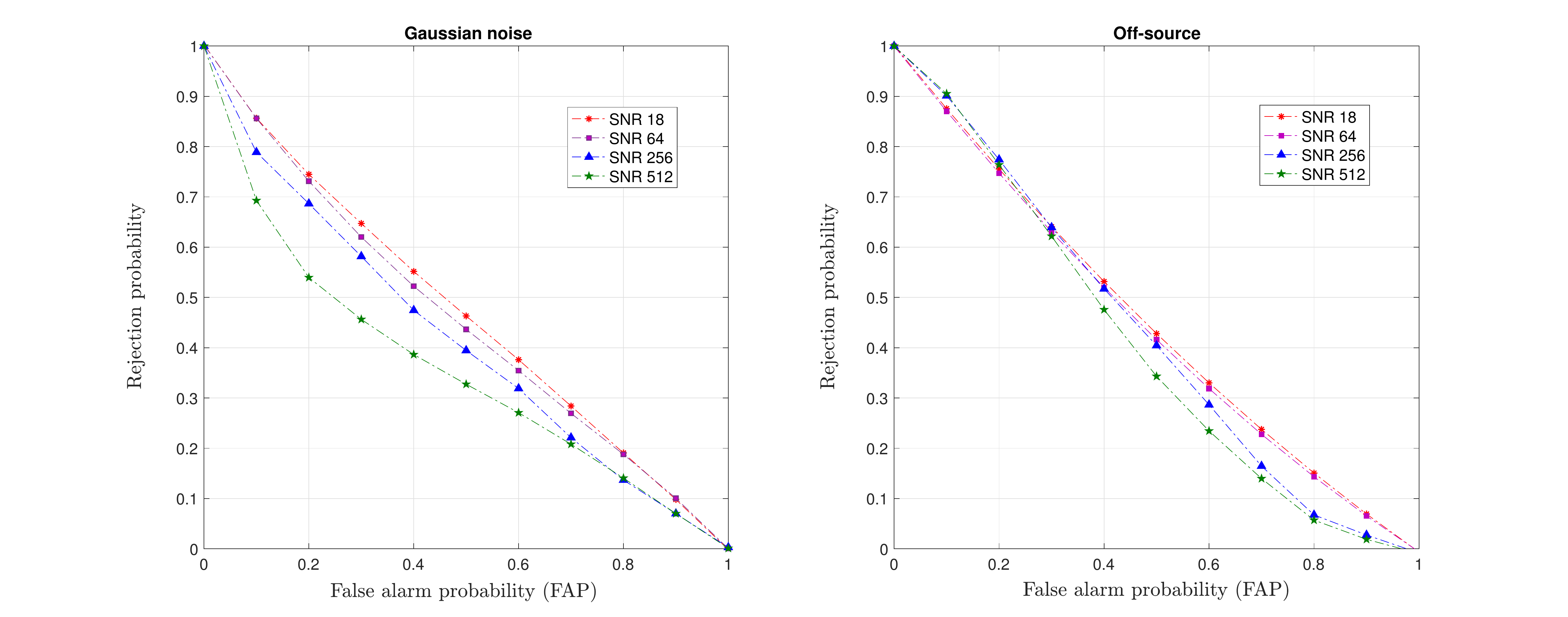}
\caption{Rejection probability of an outlier arising from persistent monochromatic noise lines in raw data. Significance for selection estimated using a Gaussian model for the noise (left) and the sky-shift method (right).
The ordinate shows the percentage of noise lines that generate an outlier with a significance greater than or equal to the threshold corresponding to the FAP in the abscissa.}
\label{fig:Rejoc}
\end{figure*}

To address this, we use as proxy simulated monochromatic noise lines at frequencies close to the targeted pulsar frequency. The putative source for wich we were looking for was again a pulsar with the rotational parameters of P03 and the sky position $\alpha=22^{\rm h}35^{\rm m}40.73^{\rm s}$, $\delta=39\degree40'44.76''$.
In particular, we produce 300 datasets with noise lines added to 4 months of Gaussian noise with varying SNRs.
Each noise line is injected with a frequency within 0.01 Hz of the targeted frequency, ensuring that the \romer{} correction will cause it to contaminate the on-source analysis. This is because, for a putative source at $f = 108.85$ Hz ((see Table \ref{tab:pulsars}), the \romer{} frequency shift will be at most \cite{2017CQGra..34m5007M}
\beq
 \Delta f= f \cdot 10^{-4} \approx 0.01 ~\mathrm{Hz}\, .
\eeq
We pick the specific frequency and phase of the noise lines from a uniform distribution.
After doing this, for every dataset we compute the significance of the noise-line outlier using sky-shifting through the 5-vector method (\sect{searches_5vector}.
As above, we produce $10^4$ background realizations using sky patches separated with a minimal distance of $0.01$ deg.
For comparison with the traditional method, we additionally evaluate the significance using the theoretical formula assuming pure Gaussian noise (cf.~\fig{5-vec_Gaussian}) \cite{Astone2010}.

We evaluate our method's ability to identify noise artifacts by studying the rejection probability as a function of the confidence threshold set to claim a detection---that is, how likely the analysis is to reject the artifact as we decrease our tolerance for false alarms (FAP).
Figure \ref{fig:Rejoc} shows the results obtained empirically with sky-shifting (right panel) and analytically assuming Gaussian noise (left panel).
In the ideal case, we would be able to perfectly measure the significance of an outlier and the curves in \fig{Rejoc} would simply be a straight line with slope $-1$.
 As we can see, strong ($\text{SNR}>256$) noise lines produce significant outliers if we assume the background to be Gaussian; however, this is not true for sky-shifting.

These tests can be extended to a general noise background.
In principle, we can model coherent instrumental noise as a linear combination of monochromatic noise lines like those injected above.
Every noise line will couple constructively or destructively with the other noise lines after Doppler corrections.
If the lines combined constructively, we would obtain a case very similar to the one presented in \fig{Rejoc} but with a larger noise-line SNR.
On the other hand, the noise lines combine destructively, then the SNR would be lower than \fig{Rejoc}.
The general case should lie somewhere in between.

\section{Conclusion} \label{sec:conclusion}

Off-sourcing can provide a much-needed efficient and robust way to empirically estimate the background of searches for continuous gravitational waves targeted at known pulsars, or as followup to all-sky searches, enabling estimates of detection significance that are valid in actual (non-Gaussian) instrumental noise.
This method has already been put into practice for diagnosing outliers in actual LIGO and Virgo searches \cite{o1cw,Abbott:2017pqa}, but a systematically study of its performance was lacking from the literature.
In this paper, we have filled in this gap by introducing the rationale behind this strategy, exploring its theoretical applicability and studying its performance in real and simulated data.

The procedure is simple: the original gravitational-wave data are time-corrected for multiple shifted sky locations to obtain as many instantiations of noise-only data, which are then analyzed by any of the usual searches with the same settings as the on-source search (\sect{offsourcing}).
Under the right conditions, we show that the sky-shifted data are blind to astrophysical signals while retaining the statistical properties of the noise.
This allows for the direct empirical estimation of the background distribution of the different search statistics.

Two conditions need to be satisfied for sky-shifting to be effective: shifted sky locations must (i) be resolvably different and (ii) be drawn from the same hemisphere as the source.
As long as this is true, sky-shifting will provide independent draws from the background distribution (\sect{blinding}).
Furthermore, for realistic signal amplitudes, the distribution will be uncontaminated by the presence of a signal at the true on-source sky location.
This is not true for extremely loud signals, but this is not a problem because in those cases the on-source statistic is always louder than the background (\sect{contamination}).
The phenomenon is analogous to that observed with strong signals in searches for compact binaries \cite{o1bbh, TheLIGOScientific:2016qqj}.

We illustrate the efficacy of sky-shifting with several examples in real and synthetic data (\sect{results_cases}).
This includes simulated Gaussian noise in the absence of signal (\fig{5-vec_Gaussian}), as well as in the presence of realistic ($\text{network-SNR}=5$, \fig{crab_medium}) and strong ($\text{network-SNR}=700$, \fig{crab_strong}) signals.
We also demonstrate the method in the presence of real LIGO O1 noise with data prepared for PSR J1932+17 \cite{o1cw} and loud hardware injection \cite{Biwer:2016oyg}.
Source parameters for all these case studies are summarized in Table \ref{tab:pulsars}.

Finally, we systematically study the performance of sky-shifting by looking at false-dismissal and false-alarm rates (\sect{results_comparison}).
The former is quantified by the receiver operating curve of \fig{RR} and the latter by the rejection-probability vs confidence-level plot of \fig{Rejoc}.
In the cases we considered, we find that sky-shifting outperforms the standard method for computing significances in the context of the 5-vector search.

\begin{acknowledgments}
We thank Jonah Kanner and Alan J.\ Weinstein for insightful comments and suggestions regarding the method; we thank Thomas Dent for useful feedback, including suggesting the ``sky-shift'' nomenclature; we also thank Graham Woan for reviewing this manuscript, as well as colleagues in the LIGO Scientific Collaboration Continuous Waves group for useful discussions.
M.I.\ is supported by NASA through the NASA Hubble Fellowship
grant No.\ HST-HF2-51410.001-A awarded by the Space Telescope
Science Institute, which is operated by the Association of Universities
for Research in Astronomy, Inc., for NASA, under contract NAS5-26555.
LIGO was constructed by the California Institute of Technology and 
Massachusetts Institute of Technology with funding from the National Science 
Foundation and operates under cooperative agreement PHY-0757058. 
We are grateful for computational resources provided by Cardiff University, and 
funded by an STFC grant supporting UK Involvement in the Operation of Advanced 
LIGO. 
For the period of this work MP was funded by the STFC under grant number {ST/N005422/1}. 
SM thanks the LIGO Visitor programme.
Plots produced using \texttt{Matplotlib} \cite{Hunter2007}. 
This paper carries LIGO Document Number \dcc. 
\end{acknowledgments}

\bibliography{gw,signals}

\end{document}